\documentclass[11pt]{article}
\usepackage{graphicx}
\usepackage{longtable,lscape}
\usepackage{amsthm}
\usepackage{amsfonts}
\usepackage{amsmath}
\usepackage{amscd}
\usepackage{mathrsfs}
\usepackage{latexsym}
\usepackage{amssymb}
\usepackage{bbm}
\usepackage{float}
\usepackage{url}

\newcommand{\captionfonts}{\footnotesize}
\makeatletter  
\long\def\@makecaption#1#2{%
  \vskip\abovecaptionskip
  \sbox\@tempboxa{{\captionfonts #1: #2}}%
  \ifdim \wd\@tempboxa >\hsize
    {\captionfonts #1: #2\par}
  \else
    \hbox to\hsize{\hfil\box\@tempboxa\hfil}%
  \fi
  \vskip\belowcaptionskip}
\makeatother 

\title{A Proposal to Extend Expected Utility in a Quantum Probabilistic Framework}
\author{Diederik Aerts\footnote{Center Leo Apostel for Interdisciplinary Studies, Krijgskundestraat 33, 1160 Brussels (Belgium). Email address: \emph{diraerts@vub.ac.be}} \quad Emmanuel Haven\footnote{School of Business and Institute IQSCS, University Road, LE1 7RH Leicester (United Kingdom). Email address: \emph{eh76@le.ac.uk}} \quad Sandro Sozzo\footnote{School of Business and Institute IQSCS, University Road, LE1 7RH Leicester (United Kingdom). Email address: \emph{ss831@le.ac.uk}}}

\date{}

\begin{document}

\maketitle

\begin{abstract}
\noindent
Expected utility theory (EUT) is widely used in economic theory. However, its subjective probability formulation, first elaborated by Savage, is linked to Ellsberg-like paradoxes and ambiguity aversion. This has led various scholars to work out non-Bayesian extensions of EUT which cope with its paradoxes and incorporate attitudes toward ambiguity. A variant of the Ellsberg paradox, recently proposed by Mark Machina and confirmed experimentally, challenges existing non-Bayesian models of decision-making under uncertainty. Relying on a decade of research which has successfully applied the formalism of quantum theory to model cognitive entities and fallacies of human reasoning, we put forward a non-Bayesian extension of EUT in which subjective probabilities are represented by quantum probabilities, while the preference relation between acts depends on the state of the situation that is the object of the decision. We show that the benefits of using the quantum theoretical framework enables the modeling of the Ellsberg and Machina paradoxes, as the representation of ambiguity and behavioral attitudes toward it. The theoretical framework presented here is a first step toward the development of a `state-dependent non-Bayesian extension of EUT' and it has potential applications in economic modeling.
\end{abstract}
\medskip
{\bf Keywords:} Expected utility theory; Ellsberg paradox; Machina paradox; quantum probability; quantum modeling.

\section{Introduction\label{intro}}
Economic theory crucially rests on the so-called `Bayesian paradigm': any source of uncertainty can be probabilistically quantified and the ensuing probability theory satisfies the axioms of Kolmogorov (Kolmogorov, 1933). Von Neumann and Morgenstern successfully applied the Bayesian paradigm when elaborating an axiomatic form of the expected utility theory (EUT), which is the predominant model of decision-making under uncertainty (von Neumann \& Morgenstern, 1944).

It is however well known that, for many events of interest, one cannot define an objective agreed-upon probability. This led Savage to extend the von Neumann and Morgenstern `objective EUT' within the Bayesian paradigm (Savage, 1954). The traditional line of reasoning is that, in the absence of objective probabilities, the decision-maker forms her/his own subjective probabilities, and takes decisions on the basis of these subjective probabilities. This is the basic tenet of subjective EUT: individuals take decisions as if they maximized expected utility with respect to a Kolmogorovian probability measure which is interpreted as their subjective probability.

The Bayesian paradigm has been applied with success to several problems in economic modeling. However, in 1961 Daniel Ellsberg demonstrated in a series of thought experiments that simple situations exist in which decision-makers violate some of the `rationality axioms' of subjective EUT, preferring certain to uncertain decisions, rather than maximizing expected utility, a behavior called by Ellsberg `ambiguity aversion' (Ellsberg, 1961). Ellsberg's experiments have been actually performed several times, generalized under different directions, and applied to domains outside decision theory, like in finance, medicine and actuarial sciences (see, e.g., McCrimmon \& Larsson, 1979; Einhorn \& Hogarth, 1986; Camerer \& Weber, 1992; Fox \& Tversky, 1995; Viscusi \& Chesson, 1999; see also the exhaustive review by Machina \& Siniscalchi, 2014). Most of the experimental findings have confirmed this ambiguity aversion attitude.

An accurate critical analysis of the problems above has led various authors to suggest that the axiomatic foundations of the Bayesian approach are not so compelling as they seem, and that the Bayesian approach is probably too limited to cope with any kind of uncertainty that affects human decision-making (Gilboa, Postlewaite \& Schmeidler, 2008). Moreover, axiomatic approaches have been elaborated which cope with the Ellsberg and other puzzles and extend EUT in a non-Bayesian sense. Relevant examples of non-Bayesian approaches to decision-making under uncertainty are, e.g., `expected utility with multiple priors' (Gilboa \& Schmeidler, 1989), `Choquet expected utility' (Schmeidler, 1989), `smooth ambiguity preferences model' (Klibanoff, Marinacci \& Mukerij, 2005), `variational preference model' (Maccheroni, Marinacci \& Rustichini, 2006), and `cumulative prospect theory' (Tversky \& Kahneman, 1992) -- just to quote the most celebrated, without the aim of being exhaustive. In addition, these non-Bayesian models have been effectively applied to economic problems of decision under uncertainty, e.g., the `home bias puzzle', the `equity premium puzzle' (see, e.g., the review by Gilboa \& Marinacci, 2013).

More recently, some scholars have proved that the decision-making models above find difficulties to represent the preferences hypothesized in two variants of the Ellsberg paradox, the `50:51 example' and the `reflection example' (Machina, 2009; Baillon, L'Haridon \& Placido, 2011). This `Machina paradox', which has been confirmed experimentally (L'Haridon \& Placido, 2010), claims for new theoretical approaches to model human preferences and decision-making in the presence of ambiguity.

We put forward in the present paper that the probabilistic formalism of quantum theory, {\it free from any physical connotation or interpretation}, can be used to model human choices in the presence of ambiguity. In this respect, we apply here the lesson we have learned from quantum physics. 
A measurement process gives rise to an interaction between the microscopic quantum entity that is measured and a measuring apparatus, and thus the latter acts as a measurement context for the former. This contextual interaction is non-controllable (which is formalized by the Heisenberg uncertainty principle), and determines an intrinsically probabilistic change of the state of the measured entity. Then, a measurement outcome is actualized among a set of possible outcomes, as a consequence of this contextual interaction. Whenever one formalizes the statistical frequencies of repeated experiments to derive probabilities, one discovers that this kind of `pure potentiality' and `contextuality' cannot be formalized in a single Kolmogorov probability space: they indeed require quantum probability (Aerts, 2009).

The quantum measurement lesson was firstly applied with success to the representation of conceptual entities. We introduced the notion of `state of a concept' to model the type of interaction that occurs when people are asked about membership, or typicality, of specific exemplars with respect to pairs of concepts and their combinations, e.g., conjunction, disjunction and negation. We applied the quantum-conceptual approach to solve various difficulties of the Kolmogorovian model of probability in conceptual combinations. Then, we extended the approach to more complex situations, showing that genuine quantum effects, i.e. `superposition', `interference', `entanglement', etc., systematically occur in the combination of natural concepts (Aerts, 2009; Aerts, Broekaert, Gabora \& Sozzo, 2013; Aerts, Gabora \& Sozzo, 2013; Aerts, Sozzo \& Veloz, 2015; Sozzo, 2015).

The quantum-conceptual approach supports a growing research that uses the mathematical formalism of quantum theory to model complex cognitive processes, like probability and similarity judgment, perception, knowledge representation and decision-making. More specifically, quantum models have shown their effectiveness in the explanation of the so-called `fallacies of human reasoning', like `conjunctive and disjunctive fallacies', `disjunction effects', `question order effects', etc. (see, e.g., Busemeyer \& Bruza, 2012; Busemeyer, Pothos, Franco \& Trueblood, 2011; Haven \& Khrennikov, 2013; Khrennikov, 2010; Pothos \& Busemeyer, 2013; Wang, Solloway, Shiffrin \& Busemeyer, 2014; Khrennikov, 2015). This novel research programme has now become a valid alternative to the Kahneman and Tversky theory of individual heuristics and biases in providing an explanation for the observed fallacies (Tversky \& Kahneman, 1974; 1983; 1992).

What about the above paradoxes of EUT? We have recently applied the quantum conceptual approach to decision-making processes. In a human decision-making process the interaction between the object of the decision and the decision-maker is of a cognitive nature rather than of a physical nature like it is the case when quantum models are applied in physics. The result of this interaction leads to the decision itself, which is actualized among the possible alternatives. Again, the probabilistic model formalizing the ambiguity occurring in the decision-making interaction cannot generally be Kolmogorovian: it should be quantum probability. In particular, the notion of state of the cognitive entity that is the object of the decision, the `decision-making (DM) entity', incorporates the notion of ambiguity and generates the (non-Kolmogorovian) quantum probability distribution modeling subjective probabilities. In addition, the change of state of the DM entity after the interaction incorporates attitudes toward ambiguity, e.g., `ambiguity aversion', or `ambiguity attraction', or even other attitudes, depending on the overall `conceptual landscape surrounding the DM entity'. We also show that preferences between acts are not absolute notions, but they depend on the state of the DM entity and on how the state of the DM changes as a consequence of the interaction with the decision-maker. This suggests associating the DM entity with a family of subjective probability distributions, parametrized by the states of the DM entity itself. We thus put forward a `state-dependent extension of EUT', of which the present formalism is a first step.

We may say that the use of quantum probabilities helps clarifying how subjective probabilities are formed. We indeed suggest that the kind of uncertainty that occurs in an ambiguity situation is similar to the kind of uncertainty that occurs in quantum theory as an effect of superposition. It is important to mention that the quantum theoretical approach allows reduction of ambiguity to risk, as subjective probabilities can be estimated from quantum probabilities. However, this is a context-dependent risk which cannot be modeled by Kolmogorovian probability. Rather, it is a `contextual risk' that is modeled in the mathematical formalism of quantum theory.

We put forward a non-Bayesian generalization of the Bayesian paradigm, where human decision-making under uncertainty is modeled in the mathematical formalism of quantum theory, quantum states incorporate the uncertainty that is present in situations of ambiguity, and state transformations incorporate people's attitude towards ambiguity. This quantum-conceptual approach was recently employed with success to model Ellsberg- and Machina-like preferences in the corresponding paradox situations, and to faithfully represent data collected in experimental tests on the `three-color Ellsberg urn' and the `Machina reflection example'.

Let us summarize the content of this paper as follows.

In Section \ref{quantumcognition} we give an overview of the technical details of the mathematical formalism of quantum theory that are needed to model cognitive 
entities, like `concepts', `conceptual combinations', or more complex `DM entities'. We then apply in Section \ref{quantumcognition} the quantum conceptual approach to model the `disjunction effect', which entails a violation of one of the axioms of subjective EUT, the `sure thing principle' (Tversky \& Shafir, 1992). Successively, we introduce in Section \ref{ellsbergmachina} the foundations of subjective EUT and briefly review its non-Bayesian extensions. We explicitly present the Ellsberg paradox in the `three-color example' and the Machina paradox in the `reflection example'. In Section \ref{quantummodels}  we instead work out a general quantum theoretical framework to represent events, states, measurements, subjective probabilities, acts and decisions, which we apply to model the Ellsberg paradox (Section \ref{quantumellsberg}) and the Machina paradox (\ref{quantummachina}). In Sections \ref{quantumellsberg} and \ref{quantummachina} we also show that the quantum theoretical framework enables a faithful representation of concrete experiments on the Ellbserg three-color example and the Machina reflection example, respectively. We finally discuss in Section \ref{conclusions} some potential applications of the quantum theoretical approach to model ambiguity and ambiguity aversion in economics and finance.

Before proceeding further, we would like to premise that the mathematical framework presented in Section \ref{quantummodels} is not derived from a representation theorem. It rather emerges from heuristic operational considerations on events, measurements, outcomes and probabilities and their representation within the formalism of quantum theory. And, indeed, it is able to reconstruct, via the quantum state of the DM entity, the subjective probability distributions that are actualized in concrete experiments as a consequence of a particular attitude toward ambiguity.

\section{Fundamentals of quantum theoretical modeling \label{quantummathematics}}
We illustrate in this section how the mathematical formalism of quantum theory can be applied to model situations outside the microscopic quantum world, more specifically, in the representation of cognitive entities. This formalism will be applied to conceptual entities in Section \ref{quantumcognition} and to decision-making (DM) entities in Section \ref{quantummodels}. We avoid in our presentation superfluous technicalities, but aim to be synthetic and rigorous at the same time.

When the quantum mechanical formalism is applied for modeling purposes, each considered entity  -- in our case a cognitive entity -- is associated with a complex Hilbert space ${\cal H}$, that is, a vector space over the field ${\mathbb C}$ of complex numbers, equipped with an inner product $\langle \cdot |  \cdot \rangle$ that maps two vectors $\langle A|$ and $|B\rangle$ onto a complex number $\langle A|B\rangle$. We denote vectors by using the bra-ket notation introduced by Paul Adrien Dirac, one of the pioneers of quantum theory (Dirac, 1958). Vectors can be `kets', denoted by $\left| A \right\rangle $, $\left| B \right\rangle$, or `bras', denoted by $\left\langle A \right|$, $\left\langle B \right|$. The inner product between the ket vectors $|A\rangle$ and $|B\rangle$, or the bra-vectors $\langle A|$ and $\langle B|$, is realized by juxtaposing the bra vector $\langle A|$ and the ket vector $|B\rangle$, and $\langle A|B\rangle$ is also called a `bra-ket', and it satisfies the following properties:

(i) $\langle A |  A \rangle \ge 0$;

(ii) $\langle A |  B \rangle=\langle B |  A \rangle^{*}$, where $\langle B |  A \rangle^{*}$ is the complex conjugate of $\langle A |  B \rangle$;

(iii) $\langle A |(z|B\rangle+t|C\rangle)=z\langle A |  B \rangle+t \langle A |  C \rangle $, for $z, t \in {\mathbb C}$,
where the sum vector $z|B\rangle+t|C\rangle$ is called a `superposition' of vectors $|B\rangle$ and $|C\rangle$ in the quantum jargon.

From (ii) and (iii) follows that the inner product $\langle \cdot |  \cdot \rangle$ is linear in the ket and anti-linear in the bra, i.e. $(z\langle A|+t\langle B|)|C\rangle=z^{*}\langle A | C\rangle+t^{*}\langle B|C \rangle$.

We recall that the `absolute value' of a complex number is defined as the square root of the product of this complex number times its complex conjugate, that is, $|z|=\sqrt{z^{*}z}$. Moreover, a complex number $z$ can either be decomposed into its cartesian form $z=x+iy$, or into its polar form $z=|z|e^{i\theta}=|z|(\cos\theta+i\sin\theta)$.  As a consequence, we have $|\langle A| B\rangle|=\sqrt{\langle A|B\rangle\langle B|A\rangle}$. We define the `length' of a ket (bra) vector $|A\rangle$ ($\langle A|$) as $|| |A\rangle ||=||\langle A |||=\sqrt{\langle A |A\rangle}$. A vector of unitary length is called a `unit vector'. We say that the ket vectors $|A\rangle$ and $|B\rangle$ are `orthogonal' and write $|A\rangle \perp |B\rangle$ if $\langle A|B\rangle=0$.

We have now introduced the necessary mathematics to state the first modeling rule of quantum theory, as follows.

\medskip
\noindent{\it First quantum modeling rule:} A state $A$ of an entity -- in our case a cognitive entity -- modeled by quantum theory is represented by a ket vector $|A\rangle$ with length 1, that is $\langle A|A\rangle=1$.

\medskip
\noindent
An orthogonal projection $M$ is a linear operator on the Hilbert space, that is, a mapping $M: {\cal H} \rightarrow {\cal H}, |A\rangle \mapsto M|A\rangle$ which is Hermitian and idempotent. The latter means that, for every $|A\rangle, |B\rangle \in {\cal H}$ and $z, t \in {\mathbb C}$, we have:

(i) $M(z|A\rangle+t|B\rangle)=zM|A\rangle+tM|B\rangle$ (linearity);

(ii) $\langle A|M|B\rangle=\langle B|M|A\rangle^{*}$ (hermiticity);

(iii) $M \cdot M=M$ (idempotency).

The identity operator $\mathbbmss{1}$ maps each vector onto itself and is a trivial orthogonal projection operator. We say that two orthogonal projections $M_k$ and $M_l$ are orthogonal operators if each vector contained in the range $M_k({\cal H})$ is orthogonal to each vector contained in the range $M_l({\cal H})$, and we write $M_k \perp M_l$, in this case. The orthogonality of the projection operators $M_{k}$ and $M_{l}$ can also be expressed by $M_{k}M_{l}=0$, where $0$ is the null operator. A set of orthogonal projection operators $\{M_k \ | \ k=1,\ldots,n\}$ is called a `spectral family' if all projectors are mutually orthogonal, that is, $M_k \perp M_l$ for $k \not= l$, and their sum is the identity, that is, $\sum_{k=1}^nM_k=\mathbbmss{1}$. A spectral family $\{M_k \ | \ k=1,\ldots,n\}$ identifies an Hermitian operator $\hat{O}=\sum_{i=1}^{n}o_kM_k$, where $o_k$ is called `eigenvalue of $\hat{O}$', i.e. is a solution of the equation $\hat{O}|o\rangle=o_k|o\rangle$ -- the non-null vectors satisfying this equation are called `eigenvectors of $\hat{O}$'.

The above definitions give us the necessary mathematics to state the second modeling rule of quantum theory, as follows.

\medskip
\noindent
{\it Second quantum modeling rule:} A measurable quantity $Q$ of an entity -- in our case a cognitive entity -- modeled by quantum theory, and having a set of possible real values $\{q_1, \ldots, q_n\}$ is represented by a spectral family $\{M_k \ | \ k=1, \ldots, n\}$, equivalently, by the Hermitian operator $\hat{Q}=\sum_{k=1}^{n}q_kM_k$, in the following way. If the conceptual entity is in a state represented by the vector $|A\rangle$, then the probability of obtaining the value $q_k$ in a measurement of the measurable quantity $Q$ is $\langle A|M_k|A\rangle=||M_k |A\rangle||^{2}$. This formula is called the `Born rule' in the quantum jargon. Moreover, if the value $q_k$ is actually obtained in the measurement, then the initial state is changed into a state represented by the vector
\begin{equation}
|A_k\rangle=\frac{M_k|A\rangle}{||M_k|A\rangle||}
\end{equation}
This change of state is called `collapse' in the quantum jargon.

\medskip
\noindent
Let us now come to the formalization of quantum probability. A major structural difference between classical probability theory, which satisfies the axioms of Kolmogorov, and quantum probability theory, which is non-Kolmogorovian, relies on the fact that the former is defined on a Boolean $\sigma$-algebra of events, whilst the latter is defined on a more general algebraic structure. More specifically, let us denote by ${\mathscr L}({\cal H})$ the set of all orthogonal projection operators over the complex Hilbert space  ${\cal H}$. ${\mathscr L}({\cal H})$ has the algebraic properties of a complete orthocomplemented lattice, but ${\mathscr L}({\cal H})$  is not distributive, hence ${\mathscr L}({\cal H})$ does not form a $\sigma$-algebra. A `generalized probability measure' over  ${\mathscr L}({\cal H})$ is a function $\mu: M \in {\mathscr L}({\cal H}) \longmapsto \mu(M) \in [0,1]$, such that $\mu(\mathbbmss{1})=1$, and $\mu(\sum_{k=1}^{\infty}M_k)=\sum_{k=1}^{\infty}\mu(M_k)$, for any countable sequence $\{ M_k \in {\mathscr L}({\cal H}) \ | \  k=1,2,\ldots \}$ of mutually orthogonal projection operators. The elements of ${\mathscr L}({\cal H})$ are said to represent `events', in this framework. Referring to the definitions above, the event ``a measurement of the quantity $Q$ gives the outcome $q_k$'' is represented by the orthogonal projection operator $M_k$.

The Born rule establishes a connection between states and generalized probability measures, as follows. 

Given a state of a cognitive entity represented by the vector $|A\rangle \in \cal H$ with length 1, it is possible to associate  $|A\rangle$ with a generalized probability measure $\mu_{A}$ over ${\mathscr L}({\cal H})$, such that, for every $M \in {\mathscr L}({\cal H})$, $\mu_{A}(M)=\langle A |M|A\rangle$. The generalized probability measure $\mu_{A}$ is a `quantum probability measure' over ${\mathscr L}({\cal H})$. Interestingly enough, if the dimension of the Hilbert space is greater than 3, all generalized probability measures over ${\mathscr L}({\cal H})$ can be written as functions $\mu_{A}(M)=\langle A|M|A\rangle$, for some unit vector $|A\rangle\in \cal H$ (Gleason theorem; Gleason, 1957).

The quantum theoretical modeling above can be extended by adding further quantum rules to model compound cognitive entities and more general classes of measurements on cognitive entities. However, the present definitions and results are sufficient to attain the results in this paper.

\section{The quantum cognition lesson\label{quantumcognition}}
In 2002 Daniel Kahneman was awarded the Nobel Prize in Economic Science for ``having integrated insights from psychological research into economic science, especially concerning human judgement and decision-making under uncertainty''.

Researchers have historically formalized human cognition and behavior by using set theoretical structures, mainly, Boolean logic and 
a probability theory that satisfies the axioms of Kolmogorov (Kolmogorov, 1933) (`Kolmogorovian probability'). These are also called `classical structures', as they were originally employed in classical physics, and later extended to economics, finance, statistics, psychology, etc. As mentioned in Section \ref{intro}, this conception has particularly flowed in economics into EUT to form the so-called `Bayesian paradigm' and, with it, the roots of `rational behavior'.

However, since the beginning of the previous century, researchers have known that the probability theory axiomatized by Kolmogorov is not the only way to talk about probabilities and uncertainty. Indeed, quantum probability formalizes uncertainty into the microscopic realm (see, e.g., (Pitowsky, 1989)). A huge amount of literature has discussed the conceptual differences between classical and quantum probability. Leaving aside these differences, that are still object of scientific debate, we can certainly pin point the structural differences between classical and quantum probabilistic theories. The former probability indeed rests a normalized probability measure on a $\sigma$-algebra of events represented by sets and set theoretical operations. The latter probability rests instead on a more abstract generalized measure on a lattice of orthogonal projection operators over a complex Hilbert space (see Section \ref{quantummathematics}).

What has become evident in the last three decades is that accumulating paradoxical findings in cognitive psychology suggest that this classical conception of logic and probability theory is also fundamentally problematical. Puzzling cognitive phenomena have been identified, revealing the so-called `fallacies of human reasoning', which can be roughly divided in two main classes, as follows.

(i) `Probability judgment errors'. People estimate the conjunction event `$A$ and $B$' (disjunction event `$A$ or $B$') as more (less) likely than the events $A$ or/and $B$ separately, which entails a violation of the monotonicity law of Kolmogorovian probability.

(ii) `Decision-making errors'. People prefer action $A$ over action $B$ if they know that an event $E$ occurs, and also if they know that $E$ does not occur, but they prefer $B$ over $A$ if they do not know whether $E$ occurs or not, which entails a violation of the total law of Kolmogorovian probability.

Fallacies of type (i) include the `conjunction' and the `disjunction fallacy', `non-monotonic reasoning' and `over/under-extension effects' in membership judgements in conceptual combinations, and violations of distance axioms in similarity judgments. Fallacies of type (ii) include the `disjunction effect' and violations of the axioms of EUT (see, e.g, Busemeyer \& Bruza (2012)), which we will extensively discuss in Section \ref{ellsbergmachina}.

While a well established proposal of solution comes from the research programme on `individual heuristics and bias' developed by Kahneman and Tversky (1979, 1983, 1992), an alternative proposal has recently grown which uses the mathematical formalism of quantum theory to model the observed deviations from classicality in human reasoning. In particular, quantum probabilistic models have shown distinctive advantages over Bayesian models in representing experimental data on the fallacies above, also allowing to make predictions and to discover new non-classical effects (see, e.g., the monographs Khrennikov, 2010; Busemeyer \& Bruza, 2012; Haven \& Khrennikov, 2013).

The origins of our quantum theoretical approach can be traced back to the studies on the structural connections between cognitive and microphysical entities, namely, their behavior with respect to `contextuality' and `pure potentiality'. We recognized that any decision process involves a `transition from potential to actual' in which an outcome is actualized from a set of possible outcomes as a consequence of a contextual interaction (of a cognitive nature) between the decision-maker and the cognitive situation that is the object of the decision. Hence, {\it human decision processes exhibit deep analogies with what occurs in a quantum measurement process, where the measurement context (of a physical nature) influences the measured quantum particle in a non-deterministic way. Quantum probability - which is able to formalize this `contextually driven actualization of potential', rather than classical probability, which only formalizes a lack of knowledge about actuality - can conceptually and mathematically cope with this situation underlying both quantum and cognitive realms} (Aerts, 2009).

The above analysis was the starting point for the development of a quantum theoretical perspective to represent conceptual entities and their combinations -- conjunctions, disjunctions and negations (Aerts, 2009; Aerts, Gabora \& Sozzo, 2013; Aerts, Broekaert, Gabora \& Sozzo, 2013; Sozzo, 2015; Aerts, Sozzo \& Veloz, 2015). We modeled different sets of experimental data that exhibited deviations from classical logical and Kolmogorovian structures, also identifying new non-classical mechanisms and extending the approach to more complex decision-making situations (Sozzo, 2015; Aerts \& Sozzo, 2016). A systematic treatment of these results would lead us beyond the scope of the present paper. Hence, we limit ourselves to summarize here a quantum theoretical modeling of the disjunction effect as an example of a long standing cognitive puzzle. The solution offered here is paradigmatic, because the disjunction effect entails a violation of the `sure thing principle', exactly like the paradoxes of subjective EUT mentioned in Section \ref{intro}. 

Savage stated the sure thing principle by means of the following story.

``A businessman contemplates buying a certain piece of property. He considers the outcome of the next presidential election relevant. So, to clarify the matter to himself, he asks whether he would buy if he knew that the Democratic candidate were going to win, and decides that he would. Similarly, he considers whether he would buy if he knew that the Republican candidate were going to win, and again finds that he would. Seeing that he would buy in either event, he decides that he should buy, even though he does not know which event obtains, or will obtain, as we would ordinarily say.'' (Savage, 1954).

Tversky and Shafir tested the sure thing principle in an experiment where they presented a group of students with a `two-stage gamble', that is, a gamble which can be played twice (Tversky \& Shafir, 1992; Busemeyer \& Bruza, 2012). At each stage the decision consisted in whether or not playing a gamble that has an equal chance of winning, say $\$200$, or losing, say $\$100$. The key result is based on the decision for the second bet, after finishing the first bet. The experiment included three situations: (i) the students were informed that they had already won the first gamble; (ii) the students were informed that they had lost the first gamble; (iii) the students did not know the outcome of the first gamble. Tversky and Shafir found that 69\%, i.e. the majority, of the students who knew they had won the first gamble chose to play again, 59\%, i.e. the majority, of the students who knew they had lost the first gamble, chose to play again; but only 36\% of the students who did not know whether they had won or lost chose to play again (equivalently, 64\%, i.e. the majority, decided not to play in the second gamble). 

The two-stage gamble experiment violates Savage's sure thing principle: students generally prefer to play again if they know they won, and they also prefer to play again if they know they lost, but they generally prefer not to play again when
they do not know whether they won or lost. More generally, the experiment performed by Tversky and Shafir violates the total law of Kolmogorovian probability. If we denote by $\mu(P)$ the total probability that a student decides to play again without knowing whether he/she has won or lost in the first gamble, by $\mu(W)$ and $\mu(L)$ the probability that the student wins or loses, respectively, by $\mu(P|W)$ the conditional probability that the student decides to play again when he/she knows he/she has won, and  by $\mu(P|L)$ the conditional probability that the student decides to play again when he/she knows he/she has lost, then it is not possible to find any value of $\mu(W)$ and $\mu(L)=1-\mu(W)$ such that $\mu(P|W)=0.69$ and $\mu(P|L)=0.59$, $p(P)=0.36$ and the law of total probability
\begin{equation}
\mu(P)=\mu(W)\mu(P|W)+\mu(L)\mu(P|L)
\end{equation}
is satisfied. This violation of the laws of Kolmogorovian probability is called the `disjunction effect'. 

An equivalent formulation of the disjunction effect is known as the `Hawaii problem', and it is again due to Tversky and Shafir (1992). Consider the following situations.

`Disjunctive version'. Imagine that you have just taken a tough qualifying examination. It is the end of the fall quarter, you feel tired and run-down, and you are not sure that you passed the exam. In case you failed you have to take the exam again in a couple of months after the Christmas holidays. You now have an opportunity to buy a very attractive 5-day Christmas vacation package to Hawaii at an exceptionally low price. The special offer expires tomorrow, while the exam grade will not be available until the following day. Would you: $x$ buy the vacation package; $y$ not buy the vacation package; $z$ pay a \$5 non-refundable fee in order to retain the rights to buy the vacation package at the same exceptional price the day after tomorrow after you find out whether or not you passed the exam?

`Pass/fail version'. Imagine that you have just taken a tough qualifying examination. It is the end of the fall quarter, you feel tired and run-down, and you find out that you passed the exam (failed the exam. You will have to take it again in a couple of months after the Christmas holidays). You now have an opportunity to buy a very attractive 5-day Christmas vacation package to Hawaii at an exceptionally low price. The special offer expires tomorrow. Would you: $x$ buy the vacation package; $y$ not buy the vacation package: $z$ pay a \$5 non-refundable fee in order to retain the rights to buy the vacation package at the same exceptional price the day after tomorrow.

In the Hawaii experiment, Tversky and Shafir experienced the same pattern of the two-stage gamble situation. Indeed, more than half of the subjects chose option $x$ (buy the vacation package) if they knew the outcome of the exam (54\% in the pass condition and 57\% in the fail condition), whereas only 32\% chose option $x$ (buy the vacation package) if they did not know the outcome of the exam. The Hawaii problem clearly shows a violation of the sure thing principle: subjects generally prefer option $x$ (buy the vacation package) when they know that they passed the exam, and they also prefer $x$ when they know that they failed the exam, but they refuse $x$ (or prefer $z$) when they don't know whether they passed or failed the exam. Moreover, as in the two-stage gamble experiment, the Hawaii problem also violates the total law of Komogorovian probability. 

A seemingly plausible explanation is that the origin of the violation of the sure thing principle in the disjunction effect is `uncertainty aversion', that is, people prefer to buy the vacation package in both cases where they have certainty about the outcome of the exam, while they refuse to buy the package when they do not yet know whether they passed or failed the exam and hence lack this certainty. We will come back to this when studying `ambiguity aversion' in Section \ref{ellsbergmachina}.

We now work out a quantum theoretical model for these two experiments, where the above mentioned deviation is described in terms of genuine quantum effects.

Let us firstly consider the Hawaii problem and denote by $A$ the conceptual situation in which the participant has passed the exam, and by $B$ the conceptual situation in which the participant has failed the exam. The disjunction of both conceptual situations, denoted by `$A$ or $B$', is the conceptual situation in which the participant `has passed or failed the exam'. The participant has to make a decision whether to buy the vacation package -- positive outcome, or not to buy it -- negative outcome.

We introduce the notion of state of a conceptual entity, as in Section \ref{quantummathematics}. Thus, each conceptual situation above is described by a defined state and represented by a unit vector of a complex Hilbert space. More explicitly, we represent $A$ by the unit vector $|A\rangle $ and $B$ by the unit vector $|B\rangle $ in a complex Hilbert space. We assume that $|A\rangle $ and $ |B\rangle $ are orthogonal, that is, $\langle A|B\rangle =0$, and represent the disjunction `$A$ or $B$' by means of the normalized superposition state vector $\frac{1} { \sqrt{2}}(|A\rangle +|B\rangle )$. The decision to be made is `to buy the vacation package' or `not to buy the vacation package'. This decision is represented by an orthogonal projection operator $M$ of the Hilbert space ${\cal H}$ in our modeling scheme. The probability of the outcome `yes', i.e. `buy the package', in the `pass' situation, i.e. state vector $|A\rangle$, is 0.54, and we denote it by $\mu(A) = 0.54$. The probability of the outcome `yes', i.e. buy the package, in the `fail' situation, i.e. state vector $|B\rangle$, is 0.57, and we denote it by $\mu(B) = 0.57$. The probability of the outcome `yes', i.e. buy the package, in the `pass or fail' situation, i.e. state vector $\frac{1}{ \sqrt{2}}(|A\rangle +|B\rangle )$, is 0.32, and we denote it by $\mu(A\ {\rm or}\ B) = 0.32$. 

In accordance with quantum probability rules in Section \ref{quantummathematics}, we have 
\begin{eqnarray}
\mu (A)&=&\langle A|M|A\rangle \\
\mu (B)&=&\langle B|M|B\rangle \\ 
\mu(A\ \mathrm{or}\ B)&=&{\frac{1}{2}}(\langle A|+\langle B|)M(|A\rangle
+|B\rangle )  \label{quantprob}
\end{eqnarray}
By applying the linearity of the Hilbert space and the hermiticity of $M$, that is, $\langle B|M|A\rangle^{\ast }=\langle A|M|B\rangle $, we then get 
\begin{eqnarray}
\mu (A\ \mathrm{or}\ B) &=&{\frac{1}{2}}(\langle A|M|A\rangle +\langle
A|M|B\rangle +\langle B|M|A\rangle +\langle B|M|B\rangle )
\nonumber \\
&=&{\frac{\mu (A)+\mu (B)}{2}}+\operatorname{Re} (\langle A|M|B\rangle) \label{interference}
\end{eqnarray}
where $\operatorname{Re}(\langle A|M|B\rangle)$ is the real part of the complex number $\langle A|M|B\rangle $, i.e. the typical interference term of quantum theory. Its presence allows to produce a deviation from the average value ${\frac{1}{2}}(\mu (A)+\mu (B))$, which would be the outcome in the absence of interference. Note that, also in this disjunction effect situation, we have applied two key quantum features, namely, `superposition', in taking ${\frac{1}{\sqrt{2}}}(|A\rangle +|B\rangle )$ to represent `$A$ or $B$', and `interference', as the effect appearing in (\ref{interference}).

Our quantum model can be realized in the three-dimensional complex Hilbert space ${\mathbb{C}}^{3}$ (Aerts, 2009; Sozzo, 2015), as follows. Let us distinguish two cases:

(i) if $\mu (A)+\mu (B)\leq 1$, we put $a(A)=1-\mu(A),$ $b(B)=1-\mu (B)$ and $\gamma =\pi$; 

(ii) if $\mu (A)+\mu (B) > 1$, we put $a(A)=\mu(A)$, $b(B)=\mu (B)$ and $\gamma =0$.

Moreover, we choose 
\begin{equation}
|A\rangle =(\sqrt{a(A)},0,\sqrt{1-a(A)})  \label{vectorA}
\end{equation}
\begin{equation}
|B\rangle=
\left\{
\begin{array}{ccc}
e^{i (\beta +\gamma )}\Big (\sqrt{\frac{(1-a(A))(1-b(B))}{a(A)}}, \sqrt{\frac{a(A)+b(B)-1}{a(A)}},-\sqrt{1-b(B)} \Big ) & {\rm if} & a(A)\ne 0 \\
e^{i\beta }(0,1,0) & {\rm if} & a(A)=0 \\
\end{array}  \label{vectorB}
\right.
\end{equation}
\begin{equation}
\beta= 
\left\{
\begin{array}{ccc}
\arccos \Big ({\frac{2\mu (A\ \mathrm{or}\ B)-\mu (A)-\mu (B)}{2\sqrt{(1-a(A))(1-b(B))}}} \Big ) & {\rm if} & a(A)\not=1,b(B)\not=1 \\
{\rm arbitrary} & {\rm if} & a(A)=1\ \mathrm{or}\ b(B)=1  \\
\end{array}  \label{anglebeta}
\right.
\end{equation}
If $\mu (A)+\mu (B)\leq 1$, we take $M$ to project orthogonally onto  the subspace of ${\mathbb{C}}^{3}$ spanned by the vector $(0,0,1)$. If $\mu (A)+\mu (B)>1$, we take $M$ to project orthogonally onto the subspace of ${\mathbb{C}}^{3}$ spanned by the 
vectors $(1,0,0)$ and $(0,1,0)$.

One can verify that this construction gives rise to a quantum mechanical representation of
the Hawaii problem situation with probabilities $\mu (A),\mu (B)$ and $\mu (A\  \mathrm{or}\ B)$. In particular, the interference term in (\ref{interference}) is given by 
\begin{equation} \label{interferenceC3}
\operatorname{Re}(\langle A|M|B\rangle) =\sqrt{(1-a(A))(1-b(B))}\cos \beta
\end{equation}
where $\beta$ is the `interference angle for the disjunction'. 

Equations (\ref{interference}) and (\ref{interferenceC3}) can be used to represent the Hawaii problem situation. If we set $\mu (A)=0.54$, $\mu (B)=0.57$ and $\mu (A\ \mathrm{or}\ B)=0.32$, and observe that $\mu (A)+\mu (B)=1.11>1$, then we have $a(A)=0.54$, $b(B)=0.57$ and $\gamma =0$. After making the calculations of (\ref{vectorA}), (\ref{vectorB}) and (\ref{anglebeta}), we obtain $|A\rangle
=(0.73,0,0.68)$, $|B\rangle =e^{i121.90^{\circ
}}(0.61,0.45,-0.66)$ and we take $M$ to project onto the subspace of ${\mathbb{C}}^{3}$ spanned by the vectors $(1,0,0)$ and $(0,1,0)$. One verifies at once that this model indeed yields the correct numerical outcomes.

Let us now come to the  two-stage gamble situation. Here, we have  $\mu (A)=0.69$, $\mu (B)=0.59$ and $\mu (A\ \mathrm{or}\ B)=0.36$, hence $\mu (A)+\mu (B)=1.28>1$, $a(A)=0.69$, $b(B)=0.59$ and $\gamma =0$. Equations (\ref{interference}) and (\ref{interferenceC3}) can be solved for $\beta=141.76^{\circ}$. In addition, (\ref{vectorA}), (\ref{vectorB}) and (\ref{anglebeta}) can be solved for $|A\rangle=(0.83,0,0.56)$, $|B\rangle =e^{i141.76^{\circ}}(0.43,0.64,-0.64)$ and $M({\mathbb{C}}^{3})$ is the subspace of ${\mathbb{C}}^{3}$ spanned by vectors $(1,0,0)$ and $(0,1,0)$. Also in this case, one easily verifies that our quantum model yields the correct numerical outcomes.

We have thus provided a quantum model which successfully represents the disjunction effect occurring in the experiments by Tversky and Shafir (1992). It is important to observe that the observed deviations from Kolmogorovian probability are not interpreted in this approach as biases of human mind  but, rather, as expressions of genuine quantum effects, namely, contextuality, emergence interference and superposition. It is also worth noticing the fundamental role that complex numbers play in our construction, since they make it possible to have a non-null interference term in (\ref{interference}).

The treatment of the disjunction effect above constitutes a relevant example of quantum modeling of cognitive entities, states, measurements, probabilities and decisions. In the next sections, we will extend this treatment to more complex decision-making situations.

\section{Expected utility theory and its pitfalls\label{ellsbergmachina}}
Researchers in probability theory distinguish between probabilities that are known, or knowable, e.g., from past data, which they call `objective probabilities', and probabilities that are not known nor can be deduced, calculated or estimated in an objective way. For this reason, Knight introduced the term `risk' to designate situations that can be described by objective probabilities, and `uncertainty' to designate situations that cannot be described by objective probabilities (Knight, 1921).\footnote{We prefer using the the term `ambiguity' when referring to situations involving unknown probabilities, as done in many textbooks and papers on the topic.} The Bayesian paradigm  mentioned in Section \ref{intro} minimizes this distinction by introducing the notion of `subjective probability': even when the probabilities are not known, people may form their own `beliefs', or `priors', thus reducing problems of decision under ambiguity to problems of decision under risk. Within the Bayesian paradigm, people then update their beliefs according to the Bayes rule of Kolmogorovian probability.
 
The predominant model of choice under risk, i.e. in the presence of objective probabilities, is the EUT elaborated by von Neumann and Morgenstern (von Neumann \& Morgenstern, 1944), which we call `objective EUT'. Von Neumann and Morgenstern presented a set of axioms allowing to uniquely represent human preferences over `lotteries' by maximization of the expected utility functional. The reasonability of their axioms is so widely accepted that these axioms constitute the `normative counterpart of rational behavior'. However, the `Allais paradox' revealed that simple situations exist where decision-makers violate some axioms of objective EUT (Allais, 1953). In addition, the von Neumann-Morgenstern framework does not apply to problems where objective probabilities are not given. For this reason, Savage extended EUT to subjective probabilities within the Bayesian paradigm above: decision-makers behave as if they had subjective probabilities with respect to which they maximize expected utility (Savage, 1954).

We illustrate in the following the main definitions and results of `subjective EUT'. While other elegant formulations of subjective EUT have been widely used in the literature, like the `Anscombe-Aumann' (Anscombe \& Aumann, 1963), or the `Fishburn' (Fishburn, 1970) formulation, we prefer presenting the Savage original formulation, as it will be naturally extended in Section \ref{quantummodels}. As in Section \ref{quantummathematics}, we try to be rigorous, without however dwelling on mathematical technicalities. 

Our basic mathematical framework requires a set ${\mathscr S}=\{\ldots, s, \ldots \}$ of states of nature. A Boolean $\sigma$-algebra of subsets of ${\mathscr S}$ is denoted by ${\mathscr A}\subseteq {\mathscr P}({\mathscr S})$ (${\mathscr P}({\mathscr S})$ is the power set of ${\mathscr S}$), while the elements of ${\mathscr A}$ denote events. A probability measure $\mu:{\mathscr A}\subseteq {\mathscr P}({\mathscr S})\longrightarrow [0,1]$ over $\mathscr A$ is such that, for every $E \in {\mathscr A}$,
\begin{equation}
\mu(E)=\int_{E}d \mu(s)
\end{equation}
Then, we denote by ${\mathscr X}$ the set of consequences. For our purposes, it is sufficient that the elements $x \in {\mathscr X}$ are monetary payoffs, so that $x$ belongs to the real line $\Re$. A decision-maker is assumed to have preferences over acts. An act is a function $f: s\in {\mathscr S} \longmapsto f(s)\in {\mathscr X}$, and we denote by ${\mathscr F}$ the set of all acts. Then, one assumes a weak order (i.e. reflexive, symmetric and transitive) relation $\succsim$ on the Cartesian product ${\mathscr F} \times {\mathscr F}$, and one introduces a utility function $u: {\mathscr X} \longrightarrow \Re$.

Savage proved that, if one assumes that the preference relation $\succsim$ satisfies a number of axioms (including the sure thing principle), then, a probability measure $\mu: {\mathscr A} \subseteq {\mathscr P}({\mathscr S})\longrightarrow [0,1]$ and a function $u:{\mathscr X}\longrightarrow \Re$ exist such that, for every $f,g \in {\mathscr F}$,
\begin{equation}\label{eut}
f \succsim g \ {\rm iff} \ \int_{S}u(f(s)) d\mu(s) \ge \int_{S}u(g(s)) d\mu(s) 
\end{equation}
The integrals in (\ref{eut}) express the expected utility functionals $W(f)$ and $W(g)$ associated with the acts $f$ and $g$, respectively. The probability measure $\mu$ is interpreted as a subjective probability measure and represents the decision-maker's beliefs, while $u$ is a utility function and represents the decision-maker's taste. In addition, $\mu$ is unique and $u$ is uniquely defined up to a positive affine transformation (see, e.g., Etner, Jeleva \& Tallon, 2012; Gilboa \& Marinacci, 2013; Karni, 2014). As in the von Neumann-Morgenstern formulation, the concavity of $u$ is a measure of the decision-maker's risk aversion (for given beliefs). Deriving both probability and utility from observed choices, Savage was able to give both a normative and descriptive status to his subjective EUT. It is however important, at this stage, to stress that the probability distribution $\mu$ is unique and satisfies the axioms of Kolmogorovian probability theory, in agreement with the Bayesian paradigm.

Subjective EUT has been widely applied to economics and finance. However, in many economic problems of interest it is not clear how one should define probabilities and, if the latter cannot be defined in a satisfactory way, how people form beliefs. In addition, in a seminal 1961 paper, Ellsberg predicted in his `two-color example' that people do not always choose by maximizing their subjective expected utility, but they generally prefer acts over events with known (or objective) probabilities to acts over events with unknown (or subjective) probabilities, a phenomenon called `ambiguity aversion' (Ellsberg, 1961). Interestingly enough, the explanation proposed by Ellsberg for the observed pattern closely resembles the `uncertainty aversion' proposed by Tversky and Shafir for the disjunction effect (see Section \ref{quantumcognition}).

The thought experiment where ambiguity aversion manifestly clashes with subjective EUT is the `three-color example'. 

Consider one urn with 30 red balls and 60 balls that are either yellow or black in unknown proportion. One ball will be drawn at random from the urn. Then, free of charge, a person is asked to bet on one of the acts $f_1$, $f_2$, $f_3$ and $f_4$ defined in Table 1. 
\noindent 
\begin{table} \label{table01}
\begin{center}
\begin{tabular}{|p{1.5cm}|p{1.5cm}|p{1.5cm}|p{1.5cm}|}
\hline
Act & Red & Yellow & Black \\ 
\hline
\hline
$f_1$ & \$100 & \$0 & \$0 \\ 
\hline
$f_2$ & \$0 & \$0 & \$100 \\ 
\hline
$f_3$ & \$100 & \$100 & \$0 \\ 
\hline
$f_4$ & \$0 & \$100 & \$100 \\ 
\hline
\end{tabular}
\end{center}
{\bf Table 1.} The payoff matrix for the Ellsberg three-color thought experiment.
\end{table}
\noindent 
Ellsberg suggested that, when asked to rank these acts, most of the persons will prefer $f_1$ over $f_2$ and $f_4$ over $f_3$. On the other hand, acts $f_1$ and $f_4$ are `unambiguous', as they are associated with events over known probabilities, while 
acts $f_2$ and $f_3$ are `ambiguous', as they are associated with events over unknown probabilities. The conclusion is that people prefer the unambiguous act over its ambiguous counterpart. There is a huge experimental evidence that decision-makers actually show this ambiguity aversion (see, e.g., the extensive review by Machina and Siniscalchi (2014)). Only the experiment by Slovic and Tversky (1974) indicated that `ambiguity attraction' was at play, while more recent experiments identify other behavioral mechanisms as primary (Binmore, Stewart \& Voorhoeve, 2012).

Neither ambiguity aversion nor ambiguity attraction can be explained within subjective EUT, as they violate the sure thing principle, according to which, preferences should be independent of the common outcome. For example, in the three-color urn, preferences should not depend on whether the common event ``a yellow ball is drawn'' pays off \$0 or \$100. More technically, subjective EUT predicts `consistency of decision-makers' preferences', that is, $f_1$ is preferred to $f_2$ if and only if $f_3$ is preferred to $f_4$. A simple calculation shows that this is impossible. Indeed, if we denote by $p_{R}$, $p_{Y}$ and $p_{B}$ the subjective probability that a red ball, a yellow ball, a black ball, respectively, is drawn (with $p_{R}=1/3=1-(p_{Y}+p_{B})$), then the expected utilities $W(f_{i})$, $i=1,2,3,4$, are such that
$W(f_1)>W(f_2)$ if and only if $(p_{R}-p_{B})(u(100)-u(0))>0$ if and only if $W(f_3)>W(f_4)$. Hence, no assignment of the subjective probabilities $p_{R}$, $p_{Y}$ and $p_{B}$ reproduces a preference with $W(f_1)>W(f_2)$  and $W(f_4)>W(f_3)$. 

In the last forty years various extensions of subjective EUT have been elaborated, mainly in an axiomatic form, to cope with the Ellsberg paradox and ambiguity aversion. These proposals weaken some of the axioms of subjective EUT, e.g., the sure thing principle, taking a non-Bayesian direction. Without pretending to be exhaustive (extensive reviews can be found in, e.g., Gilboa \& Marinacci (2013), Etner, Jelena \& Tallon (2012) and Machina \& Siniscalchi (2014)), we can roughly group these alternative decision models as follows.

(i) `Type-I models' include `Choquet expected utility' (Schmeidler, 1989) and `cumulative prospect theory' (Kahneman \& Tversky, 1979; 1992). These models assume `non-additive capacities', rather than Kolmogorovian probabilities, to represent beliefs. In particular, Choquet expected utility introduces rank-dependent axioms.

(ii) `Type-II models' include `max-min expected utility' (Gilboa \& Schmeidler, 1989) and `$\alpha$-max min expected utility` (Ghirardato, Maccheroni \& Marinacci, 2004). These models rest on the insight that, in the absence of relevant information, asking for precise subjective beliefs is too demanding. Hence, these models assume a set of probability distributions, or `multiple priors', underlying actual decisions.

(iii) `Type-III models' include `variational preference' (Maccheroni, Marinacci \& Rustichini, 2006) and `robust control' (Hansen \& Sargent, 2001). These models assume that the decision-maker has a benchmark probability distribution in mind, but he/she is not `completely confident' about it.

(iv) `Type-IV models' include `smooth ambiguity preferences' (Klibanoff, Marinacci \& Mukerij, 2005). These models still assume that the decision-maker has a set of priors, but in a concrete decision, he/she also comes up with a prior over this set of priors, or `second order belief'.

The proposals above  successfully cope with the Allais and Ellsberg paradoxes and, though some of them have been criticized (Epstein, 1999), they have manifold applications in economic and financial modeling. More important, they depart from the assumption that only Kolmogorovian probabilities can represent subjective beliefs, but they, more or less explicitly, assume non-Kolmogorovian probability distributions. 

In 2009 Mark Machina proposed new thought experiments, which seriously challenge existing results on EUT, the `50:51 example' and the `reflection example' (Machina, 2009). Like Machina and other scholars have proved, the decision models above are incompatible with the choices expected in these two examples (Machina, 2009; Baillon, l'Haridon \& Placido, 2011). In particular, the reflection example questions an axiom of Choquet expected utility, the so-called `tail separability', exactly as the Ellsberg three-color example questions the sure thing principle of subjective EUT.

We present here two versions of the reflection example, as in l'Haridon \& Placido (2010).

{\it Reflection example with lower tail shifts}.  Consider one urn with 20 balls, 10 are either red or yellow in unknown proportion, 10 are either black or green in unknown proportion. One ball will be drawn at random from the urn. Then, free of charge, a person is asked to bet on one of the acts $f_1$, $f_2$, $f_3$ and $f_4$ defined in Table 2. 
\noindent 
\begin{table} \label{table02}
\begin{center}
\begin{tabular}{|p{1.5cm}|p{1.5cm}|p{1.5cm}|p{1.5cm}|p{1.5cm}|}
\hline
Act & Red & Yellow & Black & Green \\ 
\hline
\hline
$f_1$ & \$0 & \$50 & \$25 & \$25 \\ 
\hline
$f_2$ & \$0 & \$25 & \$50 & \$25 \\ 
\hline
$f_3$ & \$25 & \$50 & \$25 & \$0 \\ 
\hline
$f_4$ & \$25 & \$25 & \$50 & \$0 \\ 
\hline
\end{tabular}
\end{center}
{\bf Table 2.} The payoff matrix for the Machina reflection example with lower tail shifts.
\end{table}
\noindent 
Machina introduced the notion of `informational symmetry', namely, the events ``the drawn ball is red or yellow'' and ``the drawn ball is black or green'' have known and equal probability and, further, the ambiguity about the distribution of
colors is similar in the two events. In an informational symmetry scenario, people should prefer act $f_1$ over act $f_2$ and act $f_4$ over act $f_3$, or they should prefer act $f_2$ over act $f_1$ and act $f_3$ over act $f_4$. On the other hand, let us introduce the utilities $u(0)$, $u(25)$ and $u(50)$, the subjective probabilities $p_R$, $p_Y$, $p_B$ and $p_G$, and calculate the expected utilities $W(f_i)$ of the acts $f_i$, $i=1,2,3,4$. Then, we have that preferences should be consistent according to subjective EUT, namely, $W(f_1)>W(f_2)$ if and only if $(u(50)-u(25))(p_Y-p_B)>0$ if and only if $W(f_3)>W(f_4)$. The interesting aspect of this example is that Choquet expected utility predicts similar consistency requirements on the basis of tail separability. 

An experiment by L'Haridon and Placido (2010) confirms the Machina preference $f_1 \succsim f_2$ and  $f_4 \succsim f_3$, consistently with informational symmetry. We will illustrate in detail the experiment in Section \ref{quantummachina}, where we will represent it within our general quantum theoretical modeling.

{\it Reflection example with upper tail shifts}.  Consider one urn with 20 balls, 10 are either red or yellow in unknown proportion, 10 are either black or green in unknown proportion. One ball will be drawn at random from the urn. Then, free of charge, a person is asked to bet on one of the acts $f_1$, $f_2$, $f_3$ and $f_4$ defined in Table 3. 
\noindent 
\begin{table} \label{table03}
\begin{center}
\begin{tabular}{|p{1.5cm}|p{1.5cm}|p{1.5cm}|p{1.5cm}|p{1.5cm}|}
\hline
Act & Red & Yellow & Black & Green \\ 
\hline
\hline
$f_1$ & \$50 & \$50 & \$25 & \$75 \\ 
\hline
$f_2$ & \$50 & \$25 & \$50 & \$75 \\ 
\hline
$f_3$ & \$75 & \$50 & \$25 & \$50 \\ 
\hline
$f_4$ & \$75 & \$25 & \$50 & \$50 \\ 
\hline
\end{tabular}
\end{center}
{\bf Table 3.} The payoff matrix for the Machina reflection example with upper tail shifts.
\end{table}
\noindent 
According to Machina's informational symmetry, people should again prefer act $f_1$ over act $f_2$ and act $f_4$ over act $f_3$, or they should prefer act $f_2$ over act $f_1$ and act $f_3$ over act $f_4$. On the other hand, let us introduce the utilities $u(25)$, $u(50)$ and $u(75)$, the subjective probabilities $p_R$, $p_Y$, $p_B$ and $p_G$, and calculate the expected utilities $W(f_i)$ of the acts $f_i$, $i=1,2,3,4$. Then, we have that preferences should again be consistent according to subjective EUT, namely, $W(f_1)>W(f_2)$ if and only if $(u(50)-u(25))(p_Y-p_B)>0$ if and only if $W(f_3)>W(f_4)$. One shows that tail separability of Choquet expected utility leads to a similar prediction. 
 
 The experiment by L'Haridon and Placido (2010) confirms again the Machina preference $f_1 \succsim f_2$ and  $f_4 \succsim f_3$, consistently with informational symmetry, and its results will be reviewed in Section \ref{quantummachina}.

The theoretical and experimental arguments above strongly require a novel approach to ambiguity. As mentioned in Section \ref{intro}, a possible way out of these difficulties is assuming non-Kolmogorovian probability distributions to represent subjective beliefs. Following our considerations in Section \ref{quantumcognition}, we believe that quantum probability distribution is a proper candidate to represent the uncertainty surrounding decision-making situations.

\section{A theoretical framework to represent preferences\label{quantummodels}}
Inspired by the  cognitive approach in Section \ref{quantumcognition}, we have recently worked out a quantum theoretical framework to model the Ellsberg and Machina paradox situations (Aerts, Sozzo \& Tapia, 2012; 2014). We have also successfully represented data collected on the three-color Ellsberg (Aerts, Sozzo \& Tapia, 2014) and Machina reflection (L'Haridon \& Placido, 2010) experiments  within the quantum framework (Aerts \& Sozzo, 2016). In this section we generalize these results by presenting a unified perspective to model events, states, subjective probabilities, acts, preferences and decisions within the quantum mechanical formalism. This can be considered as a first step toward the elaboration of `state dependent EUT' where subjective probabilities are represented by quantum probabilities, and attitudes toward ambiguity are incorporated into the states of the cognitive entities.

We begin with the introduction of some basic notions and definitions that are needed to operationally describe the cognitive aspects of decision-making under uncertainty, following the approach in Section \ref{quantumcognition}.

(i) The cognitive situation that is the object of the decision identifies a DM entity in a defined state $p_v$. We denote by $\Sigma_{DM}$ the set of all possible states of the DM entity. This state has a cognitive nature, hence it should be distinguished from a physical state or a state of nature (see Section \ref{ellsbergmachina}). The cognitive state mathematically captures aspects of ambiguity.

(ii) There is a contextual interaction of a cognitive, {\it not physical}, nature between the decision-maker and the DM entity. This contextual interaction determines a change of the state of the DM entity. The way in which this change occurs depends on subjective attitudes toward ambiguity (ambiguity aversion, ambiguity attraction, etc.).

(iii) Events correspond to measurements that can be performed on the DM entity. For each state $p_v$ of the DM entity, each event $E$ is associated with a probability $\mu_{v}(E)$ that the event occurs when the DM entity is in the state $p_v$.

(iv) The DM entity, its states, events, probabilities and the decision-making process are modeled by using the mathematical formalism of quantum theory in Section \ref{quantummathematics}. In particular, the state of the DM entity identifies a single quantum probability distribution via the Born rule. We interpret this quantum probability distribution, which is generally non-Kolmogorovian, as the subjective probability distribution associated with the specific DM process.

Let us start with the simple case where the set ${\mathscr S}$ of states of nature is finite and let $\{ E_1, E_2, \ldots, E_n \}$ be a set of mutually exclusive and exhaustive elementary events, which form a partition of ${\mathscr S}$. We denote by ${\mathscr X}$ the set of consequences, and suppose that ${\mathscr X}$ contains monetary outcomes, for the sake of simplicity. An act is defined as a function $f:{\mathscr S} \longrightarrow {\mathscr X}$, and we denote the set of acts by ${\mathscr F}$. If the act $f$ maps the elementary event $E_i$ into the outcome $x_i\in \Re$, then we can equivalently define $f$ by the 2n-tuple $(E_1,x_1;\ldots;E_n,x_n)$. We assume that a utility function $u: {\mathscr X} \longrightarrow \Re$ exists over the set of consequences which incorporates individual preferences toward risk.

We refer to the mathematics introduced in Section \ref{quantummathematics}. The DM entity is associated with a Hilbert space $\cal H$ over the field $\mathbb C$ of complex numbers. Since $n$ is the number of elementary events, the space $\cal H$ can be chosen to be isomorphic to the Hilbert space ${\mathbb C}^n$ of n-ples of complex numbers. We thus denote by $\{ |\alpha_1\rangle, |\alpha_2\rangle, \ldots, |\alpha_n\rangle\}$ the canonical orthonormal (ON) basis of ${\mathbb C}^n$, that is, $|\alpha_1\rangle=(1,0,\ldots 0)$, \ldots, $|\alpha_n\rangle=(0,0,\ldots n)$. The event $E_i$ is then represented by the orthogonal projection operator $P_i=|\alpha_i \rangle \langle \alpha_i|$, $i \in \{1,\ldots, n \}$.

For every state $p_v \in \Sigma_{DM}$ of the DM entity, represented by the unit vector $|v\rangle=\sum_{i=1}^{n}\langle \alpha_i|v\rangle |\alpha_i\rangle\in {\mathbb C}^n$, the quantum probability distribution
\begin{equation}
\mu_{v}: P \in {\mathscr L}( {\mathbb C}^n) \longmapsto  \mu_{v}(P)\in [0,1]
\end{equation}
(${\mathscr L}( {\mathbb C}^n)$ is the lattice of all orthogonal projection operators over the complex Hilbert space ${\mathbb C}^n$) induced by the Born rule, associates the probability that the event $E$, represented by the orthogonal projection operator $P$, occurs when the DM entity is in the state $p_v$. Thus, in particular,
\begin{equation} \label{quantumprobability}
\mu_{v}(E_i)=\langle v|P_i|v\rangle=|\langle \alpha_i|v\rangle|^{2}
\end{equation}
for every $i\in \{ 1, \ldots, n\}$.

Suppose that, when the decision-maker is presented with a choice between the acts $f$ and $g$, the DM entity is in the initial state $p_{v_0}$. This state is interpreted as the state of the DM entity when no cognitive context is present. As the decision-maker starts pondering between $f$ and $g$, this mental action can be described as a cognitive context interacting with the DM entity and changing its state. The type of state change directly depends on the decision-maker's attitude toward ambiguity. More precisely, if the DM entity is in the initial state $p_{v_0}$ and the decision-maker is asked to choose between the acts $f$ and $g$, a given attitude toward ambiguity, say ambiguity aversion, will determine a given change of state of the DM entity to a state $p_v$, leading the decision-maker to prefer, say $f$ to $g$. But, a different attitude toward ambiguity will determine a different change of state of the DM entity to a state $p_{w}$, leading the decision-maker to prefer $g$ to $f$. In this way, different attitudes toward ambiguity are formalized by different changes of state inducing different subjective probabilities. 

The considerations above suggest associating the DM entity with a `family of subjective probability distributions' $\{\mu_{v}:{\mathscr L}( {\mathbb C}^n) \longrightarrow [0,1]  \ | \ p_{v} \in \Sigma_{DM}  \}$, represented by quantum probabilities and parametrized by the state of the DM entity. In a concrete choice between two acts we will derive the exact state, hence the specific subjective probability distribution that represents the actual choice.

Let us now come to the representation of acts. The act $f=(E_1,x_1;\ldots;E_n,x_n)$ is represented by the Hermitian operator
\begin{equation} \label{quantumact}
\hat{F}=\sum_{i=1}^{n} u(x_i)P_i=\sum_{i=1}^{n} u(x_i)|\alpha_i\rangle\langle \alpha_i|
\end{equation}
For every $p_v \in \Sigma_{DM}$, we introduce the functional `expected utility in the state $p_v$' $W_v:{\mathscr F} \longrightarrow \Re$ as follows. 

For every $f \in {\mathscr F}$,
\begin{equation} \label{quantumexpectedutility}
W_{v}(f)=\langle v| \hat{F}| v \rangle=\langle v| \Big ( \sum_{i=1}^{n} u(x_i)P_i   \Big ) |v\rangle=\sum_{i=1}^{n} u(x_i) \langle v|P_i|v\rangle=\sum_{i=1}^{n} u(x_i) |\langle \alpha_i|v\rangle|^{2}=\sum_{i=1}^{n}u(x_i) \mu_{v}(E_i)
\end{equation}
because of (\ref{quantumprobability}) and (\ref{quantumact}). Equation (\ref{quantumexpectedutility}) generalizes the expected utility formula of subjective EUT. As we can see, expected utility explicitly depends on the state $p_v$ of the DM entity. This means that, for two acts $f$ and $g$, two states $p_v$ and $p_w$ may exist such that $W_{v}(f)>W_{v}(g)$,  but $W_{w}(f)<W_{w}(g)$, depending on subjective attitudes toward ambiguity. This suggests introducing a state-dependent preference relation $\succsim_{v}$ on the set of acts $\mathscr F$, as follows. 

For every $f,g \in {\mathscr F}$, $p_{v}\in \Sigma_{DM}$,
\begin{equation}
f \succsim_{v} g \ {\rm iff} \ W_{v}(f) \ge W_{v}(g)
\end{equation}
It follows that the DM entity incorporates the presence of ambiguity, as the quantum probability distribution representing subjective probabilities depends on the state of the DM entity. Furthermore, the way in which the state of the DM entity changes in the interaction with the decision-maker incorporates people's attitude toward ambiguity, as it determines the state-dependent preference relation $\succsim_{v}$. The state-dependence enables the `inversion of preferences' observed in the Ellsberg and Machina paradox situations, as will become evident from the next sections.

\subsection{Application to the Ellsberg three-color urn\label{quantumellsberg}}
Let us now consider the Ellsberg three-color example and represent it by using the formalism in Section \ref{quantummodels}.

The DM entity, which we call the `Ellsberg entity', is a cognitive representation of the urn with 30 red balls and 60 yellow and black balls in unknown proportion. Let  us point out explicitly that the `Ellsberg entity' is `not' the physical entity of the urn with 30 red balls and 60 yellow and black balls. We can also easily understand that it cannot be, because for the physical entity the proportion of black balls and yellow balls `is' always determined. Thus, if we specify in our description of the Ellsberg situation that this proportion is unknown we bring in explicitly a cognitive element, which makes the Ellsberg entity, i.e. the entity experimented upon, a cognitive entity and no longer a physical entity. One can wonder whether this cognitive aspect which render the Ellsberg entity from a physical entity into a cognitive entity is not just a subjective element which can be recuperated in the notion of subjective probability. It is not, at least not if subjective probabilities are meant to describe the subjectiveness pertained to the person performing the Ellsberg test. Indeed, the specification of the proportion of black balls and yellow balls to be unknown is presented each time again to each different participant when the test is performed, and hence cannot be attributed to the subjectivity of one individual test participant. It is objectively part of the beginning experimental situation that each person is confronted with in an experiment, and this is the reason it pertains to the cognitive representation of the Ellsberg physical entity which is indeed the `object' of the test. The above reasoning clarifies why we introduce the notion of `state' to describe it. Hence, the state $p_v$ of the Ellsberg entity exactly describes the cognitive situation of the urn with 30 red balls and 60 yellow and black balls in unknown proportion. However, many states of the Ellsberg entity are possible if we, for example, also specify something more than just `unknown proportion' about yellow and black balls, in which case the cognitive situation is described by a different state. We represent each state $p_v$ by the unit vector $|v\rangle$ of the complex Hilbert space ${\mathbb C}^{3}$ over complex numbers.\footnote{As mentioned in Section \ref{quantummodels}, the choice of ${\mathbb C}^{3}$ depends on the fact that there are three mutually exclusive and exhaustive events in the three-color example -- the generalization to the Ellsberg n-color example is straightforward.} We denote by $(1,0,0)$, $(0,1,0)$ and $(0,0,1)$ the unit vectors of the canonical basis of ${\mathbb C}^{3}$.

Let us now consider the elementary, exhaustive and mutually exclusive events $E_R$: ``a red ball is drawn'', $E_Y$: ``a yellow ball is drawn'', and $E_B$: ``a black ball is drawn''. They define a `color measurement' on the Ellsberg entity. This color measurement has three outcomes, corresponding to the three colors red, yellow and black, and it is represented by a Hermitian operator with eigenvectors $|R\rangle=(1,0,0)$, $|Y\rangle=(0,1,0)$,  and $|B\rangle=(0,0,1)$ or, equivalently, by the spectral family $\{ P_{i}=|i\rangle\langle i | \ | \ i=R,Y,B  \}$. In other terms, the event $E_i$ is represented by the orthogonal projection operator $P_i=|i\rangle\langle i|$, $i=R,Y,B$. In the canonical basis of ${\mathbb C}^{3}$, a state $p_v$ of the Ellsberg entity is represented by the unit vector
\begin{equation}
|v \rangle=\rho_R e^{i \theta_{R}}|R\rangle+\rho_Y e^{i \theta_{Y}}|Y\rangle+\rho_B e^{i \theta_{B}}|B\rangle=(\rho_R e^{i \theta_{R}},\rho_Y e^{i \theta_{Y}},\rho_B e^{i \theta_{B}})
\end{equation}
By using the Born rule of quantum probability, the probability $\mu_{v}(E_i)$ of drawing a ball of color $i$, $i=R,Y,B$, when the Ellsberg entity is in a state $p_v$, is given by
\begin{equation}
\mu_{v}(E_i)=\langle v | P_{i} | v \rangle=|\langle i | v \rangle|^{2}=\rho_{i}^{2}
\end{equation}
We have $\rho_{R}^{2}=1/3$, as the urn contains 30 red balls. Therefore, a state $p_v$ of the Ellsberg entity is represented by the unit vector 
\begin{equation}  \label{ellsbergstate}
|v \rangle=(\frac{1}{\sqrt{3}} e^{i \theta_{R}},\rho_Y e^{i \theta_{Y}}, \sqrt{\frac{2}{3}-\rho_{Y}^{2}} e^{i \theta_{B}})
\end{equation}
Let us now introduce two specific states $p_{RY}$ and $p_{RB}$ of the Ellsberg entity. The states $p_{RY}$ and $p_{RB}$ are represented by the unit vectors
\begin{equation} \label{noblack}
|v_{RY} \rangle=(\frac{1}{\sqrt{3}} e^{i \theta_{R}},\sqrt{\frac{2}{3}} e^{i \theta_{Y}}, 0)
\end{equation}
and 
\begin{equation} \label{noyellow}
|v_{RB} \rangle=(\frac{1}{\sqrt{3}} e^{i \theta_{R}},0, \sqrt{\frac{2}{3}} e^{i \theta_{B}})
\end{equation}
and describe the cognitive situation ``there are no black balls'' and ``there are no yellow balls'', respectively.

As the Ellsberg entity is a cognitive entity, `cognitive contexts' have an influence on its state, and in general will make a specific state change into another state. This is what happens in the cognitive realm in analogy with the physical realm, where a physical context will in general change the physical state of a physical entity. Hence, whenever a decision-maker is asked to ponder between the choice of $f_1$ and $f_2$, the pondering itself, before a choice is made, is a cognitive context, and hence it changes in general the state of the Ellsberg entity. Similarly, whenever a decision-maker is asked to ponder between the choice of $f_3$ and $f_4$, also this introduces a cognitive context, before the choice is made  -- and a context which in general is different from the one introduced by pondering about the choice between $f_1$ and $f_2$ -- which will in general change the state of the Ellsberg entity.

Let us now introduce a state $p_0$ describing the situation where no cognitive context is present. This is the initial state of the Ellsberg entity, and symmetry reasons suggest to represent it by the unit vector $|v_0\rangle=\frac{1}{\sqrt{3}}(1,1,1)$. Then, a pondering about the choice between $f_1$ and $f_2$ will make the state $p_0$ of the Ellsberg entity change to a state $p_{w_1}$ that is generally different from the state $p_{w_2}$ in which the Ellsberg entity changes from $p_0$ when pondering about a choice between $f_3$ and $f_4$. In particular, a `highly ambiguity averse' decision-maker will be, as a consequence of her/his pondering in the choice between $f_1$ and $f_2$, confronted with the Ellsberg entity which changes from the initial state $p_0$ to a state $p_{w_1}$ that is very close to the state $p_{RY}$ represented in (\ref{noblack}). Analogously, a `highly ambiguity averse' decision-maker will be, as a consequence of her/his pondering in the choice between $f_3$ and $f_4$, confronted with the Ellsberg entity which changes from the state $p_0$ to a state $p_{w_2}$ that is very close to the state $p_{RB}$ represented in (\ref{noyellow}).

Let us then come to the representation of the acts $f_1$, $f_2$, $f_3$ and $f_4$ in Table 1, Section \ref{ellsbergmachina}. For a given utility function $u$, to be estimated from empirical data, we respectively associate $f_1$, $f_2$, $f_3$ and $f_4$ with the Hermitian operators
\begin{eqnarray}
\hat{F}_{1}&=&u(100)P_R+u(0)P_Y+u(0)P_B \label{f1}\\
\hat{F}_{2}&=&u(0)P_R+u(0)P_Y+u(100)P_B \label{f2}\\
\hat{F}_{3}&=&u(100)P_R+u(100)P_Y+u(0)P_B\label{f3} \\
\hat{F}_{4}&=&u(0)P_R+u(100)P_Y+u(100)P_B \label{f4}
\end{eqnarray}
The corresponding expected utilities in a state $p_{v}$ of the Ellsberg entity are 
\begin{eqnarray}
W_{v}(f_1)&=&\langle v| \hat{F}_{1}|v\rangle=\frac{1}{3}u(100)+\frac{2}{3}u(0) \label{w1}\\
W_{v}(f_2)&=&\langle v| \hat{F}_{2}|v\rangle=(\frac{1}{3}+\rho_Y^2)u(0)+(\frac{2}{3}-\rho_{Y}^{2})u(100) \label{w2} \\
W_{v}(f_3)&=&\langle v| \hat{F}_{3}|v\rangle=(\frac{1}{3}+\rho_Y^2)u(100)+(\frac{2}{3}-\rho_{Y}^{2})u(0) \label{w3}\\
W_{v}(f_4)&=&\langle v| \hat{F}_{4}|v\rangle=\frac{1}{3}u(0)+\frac{2}{3}u(100) \label{w4}
\end{eqnarray}
We observe that $W_{v}(f_1)$ and $W_{v}(f_4)$ do not depend on the state $p_v$, hence they are ambiguity-free, i.e. independent of the state, while $W_{v}(f_2)$ and $W_{v}(f_3)$ do depend on $p_v$. This means that it is possible to find a state $p_{w_1}$, e.g., the state represented in (\ref{noblack}), such that $W_{w_1}(f_1) > W_{w_1}(f_2)$,  and a state $p_{w_2}$, e.g., the state represented in (\ref{noyellow}), such that $W_{w_2}(f_4) > W_{w_2}(f_3)$. These two states reproduce Ellsberg preferences, in agreement with an ambiguity aversion attitude.

We repeated the Ellsberg three-color experiment, asking 57 persons, chosen among our colleagues and friends, to rank the four acts in Table 1, Section \ref{ellsbergmachina} (Aerts, Sozzo \& Tapia, 2014). We found that 34 participants preferred acts $f_1$ and $f_4$, 12 participants preferred acts $f_2$ and $f_3$, 7 participants preferred acts $f_2$ and $f_4$, and 6 participants preferred acts $f_1$ and $f_3$. This makes the weights with preference of acts $f_1$ over act $f_2$ to be 0.68 against 0.32, and the weights with preference of act $f_4$ over act $f_3$ to be 0.69 against 0.31. Hence, 46 participants over 57 chose $f_1$ and $f_4$ or the inversion $f_2$ and $f_3$, for an `inversion percentage' of 78\%, thus confirming the typical behavior observed in the Ellsberg three-color example.

A quantum model for these data can be constructed by finding two orthogonal states $p_{w_1}$ and $p_{w_2}$, represented by the unit vectors $|w_1\rangle$ and $|w_2\rangle$, respectively, such that
\begin{eqnarray}
\langle w_1|\hat{F}_{1}-\hat{F}_{2}|w_1\rangle=0.68 \label{data1} \\
\langle w_2|\hat{F}_{4}-\hat{F}_{3}|w_2\rangle=0.69 \label{data2}
\end{eqnarray} 
where $\hat{F}_{i}$, $i=1,2,3,4$, are defined in (\ref{f1})--(\ref{f4}). In the canonical basis of ${\mathbb C}^{3}$, the solution is
\begin{eqnarray}
|w_1 \rangle&=&(\frac{1}{\sqrt{3}}, 0.787 e^{i 28^{\circ}}, 0.216 e^{i 9.3^{\circ}}) \label{w1ellsberg}  \\
|w_2 \rangle&=&(\frac{1}{\sqrt{3}},0.206 e^{i 208^{\circ}}, 0.790 e^{i 189.3^{\circ}}) \label{w2ellsberg}
\end{eqnarray}
as we have proved in Aerts \& Sozzo (2016). The states $p_{w_1}$ and $p_{w_2}$ represented in (\ref{w1ellsberg}) and (\ref{w2ellsberg}) identify the subjective probability distributions $\mu_{w_1}$ and $\mu_{w_2}$, respectively, reproducing the ambiguity aversion pattern of the experiment, for an utility value $u(100)-u(0)\approx 2.4$. But, generally speaking, preferences do depend on the state $p_v$ of the Ellsberg entity. This completes the construction of a quantum model for the Ellsberg three-color example, which follows the prescriptions in Section \ref{quantummodels}.

\subsection{Application to the Machina reflection example\label{quantummachina}}
In this section we represent the `Machina reflection example' within the formalism. We start from the reflection example with lower tail shifts. The DM entity, which we call the `Machina entity', is the urn with 10 red or yellow balls and 10 black or green balls, in both cases in unknown proportion. A possible state $p_v$ of the Machina entity has a cognitive nature and is represented by the unit vector $|v\rangle$ of the complex Hilbert space ${\mathbb C}^{4}$ over complex numbers. We denote by $(1,0,0,0)$, $(0,1,0,0)$, $(0,0,1,0)$ and $(0,0,0,1)$ the unit vectors of the canonical basis of ${\mathbb C}^{4}$.

Let us consider the elementary, exhaustive and mutually exclusive events $E_R$: ``a red ball is drawn'', $E_Y$: ``a yellow ball is drawn'', $E_B$: ``a black ball is drawn'', and $E_G$: ``a green ball is drawn''. They define a `color measurement' that can be performed on the Machina entity. This color measurement has four outcomes corresponding to the four colors red, yellow, black and green, and it is represented by a Hermitian operator with eigenvectors $|R\rangle=(1,0,0,0)$, $|Y\rangle=(0,1,0,0)$, $|B\rangle=(0,0,1,0)$ and $|G\rangle=(0,0,0,1)$ or, equivalently, by the  spectral family $\{ P_{i}=|i\rangle\langle i | \ | \ i=R,Y,B,G \}$. Hence, the event $E_i$ is represented by the orthogonal projection operator $P_i$, $i=R,Y,B,G$. In the canonical basis of ${\mathbb C}^{4}$, a state $p_v$ of the Machina entity is represented by the unit vector
\begin{equation}
|v \rangle=\rho_R e^{i \theta_{R}}|R\rangle+\rho_Y e^{i \theta_{Y}}|Y\rangle+\rho_B e^{i \theta_{B}}|B\rangle+\rho_G e^{i \theta_{G}}|G\rangle=(\rho_R e^{i \theta_{R}}, \rho_Y e^{i \theta_{Y}}, \rho_B e^{i \theta_{B}},  \rho_G e^{i \theta_{G}})
\end{equation}
By using quantum probabilistic rules, the probability $\mu_{v}(E_i)$ of drawing a ball of color $i$, $i=R,Y,B,G$, when the Machina entity is in a state $p_v$ is given by
\begin{equation}
\mu_{v}(E_i)=\langle v | P_{i} | v \rangle=|\langle i | v \rangle|^2=\rho_{i}^{2}
\end{equation}
The reflection with lower tail shifts situation requires that $\rho_{R}^{2}+\rho_{Y}^2=1/2=\rho_{B}^{2}+\rho_{G}^2$. Therefore, 
 a state $p_v$ of the Machina entity is represented by the unit vector 
\begin{equation} \label{machinastate}
|v \rangle=(\rho_R e^{i \theta_{R}},\sqrt{\frac{1}{2}-\rho_R^2} e^{i \theta_{Y}}, \rho_B e^{i \theta_{B}}, \sqrt{\frac{1}{2}-\rho_B^2} e^{i \theta_{G}})
\end{equation}
As in the Ellsberg case, let us suppose that the initial state $p_0$ of the Machina entity completely reflects the symmetry between the different colors. Thus, $p_0$ is represented by the unit vector $|v_0\rangle=\frac{1}{2}(1,1,1,1)$. Whenever the decision-maker is presented with the Machina paradox situation, her/his pondering about the choices to make gives rise to a cognitive context, which changes the state of the Machina entity from $p_0$ to a generally different state $p_{w}$, represented by the unit vector $|w\rangle$, as in (\ref{machinastate}). In this framework, the pondering about a choice between $f_1$ and $f_2$ will make the initial state $p_0$ of the Machina entity change to a state $p_{w_1}$ that is generally different from the state $p_{w_2}$ in which the Machina entity changes from the initial state $p_0$ in a pondering about the choice between $f_3$ and $f_4$. 

Let us now come to the representation of the acts $f_1$, $f_2$, $f_3$ and $f_4$ in Table 2, Section \ref{ellsbergmachina}. For a given utility function $u$, to be estimated from empirical data, we respectively associate $f_1$, $f_2$, $f_3$ and $f_4$ with the Hermitian operators
\begin{eqnarray}
\hat{F}_{1}&=&u(0)P_R+u(50)P_Y+u(25)P_B+u(25)P_G \label{f1mach}\\
\hat{F}_{2}&=&u(0)P_R+u(25)P_Y+u(50)P_B+u(25)P_G \label{f2mach}\\
\hat{F}_{3}&=&u(25)P_R+u(50)P_Y+u(25)P_B+u(0)P_G \label{f3mach} \\
\hat{F}_{4}&=&u(25)P_R+u(25)P_Y+u(50)P_B+u((0)P_G \label{f4mach}
\end{eqnarray}
The corresponding expected utilities in a state $p_{v}$ are 
\begin{eqnarray}
W_{v}(f_1)&=&\langle v| \hat{F}_{1}|v\rangle=u(0)\rho_R^2+u(50)\rho_Y^2+\frac{1}{2}u(25) \label{w1mach}\\
W_{v}(f_2)&=&\langle v| \hat{F}_{2}|v\rangle=u(0)\rho_R^2+u(25)\rho_Y^2+u(50)\rho_B^2+u(25)\rho_G^2 \label{w2mach} \\
W_{v}(f_3)&=&\langle v| \hat{F}_{2}|v\rangle=u(25)\rho_R^2+u(50)\rho_Y^2+u(25)\rho_B^2+u(0)\rho_G^2 \label{w3mach}\\
W_{v}(f_4)&=&\langle v| \hat{F}_{4}|v\rangle=\frac{1}{2}u(25)+u(50)\rho_B^2+u(0)\rho_G^2 \label{w4mach}
\end{eqnarray}
All expected utilities depend on the state $p_v$, thus it is possible to find a state $p_{w_1}$ such that $W_{w_1}(f_1) > W_{w_1}(f_2)$ ($W_{w_1}(f_2) > W_{w_1}(f_1)$), and a state $p_{w_2}$ such that $W_{w_2}(f_4) > W_{w_2}(f_3)$ ($W_{w_2}(f_3) > W_{w_2}(f_4)$).  Indeed, let us consider the state $p_{YG}$ describing the cognitive situation where there are no red balls and no black balls. This state is represented by the unit vector 
\begin{equation} \label{noredblackMachina}
|v_{YG} \rangle=(0,\sqrt{\frac{1}{2}} e^{i \theta_{Y}}, 0, \sqrt{\frac{1}{2}} e^{i \theta_{G}})
\end{equation}
Then, let us consider the state $p_{RB}$ describing the cognitive situation where there are no yellow balls and no green balls. This state is represented by the unit vector 
 \begin{equation} \label{noyellowgreenMachina}
|v_{RB} \rangle=(\sqrt{\frac{1}{2}} e^{i \theta_{R}},0, \sqrt{\frac{1}{2}} e^{i \theta_{B}}, 0)
\end{equation}
By using (\ref{w1mach}), (\ref{w2mach}),  (\ref{w3mach}) and (\ref{w4mach}), we have $W_{YG}(f_1)=W_{RB}(f_4)$ and  $W_{YG}(f_2)=W_{RB}(f_3)$. Therefore, the states $p_{YG}$ and $p_{RB}$ perfectly reproduce informational symmetry in the Machina reflection example with lower tail shifts.

Let us now apply the quantum model for the Machina paradox situation to represent the data collected in L'Haridon \& Placido (2010) on the reflection example with lower tail shifts. These authors asked 94 students to rank the four acts  in Table 2, Section \ref{ellsbergmachina}. The students' response was that 11 students preferred acts $f_1$ and $f_3$, 
44 students preferred acts $f_1$ and $f_4$, 
15 students preferred acts $f_2$ and $f_4$, and
24 students preferred acts $f_2$ and $f_3$. This entails that $68$ students over 94 reversed their preferences, for an inversion percentage of $72\%$, thus violating subjective EUT and in agreement with Machina's expectations (L'Haridon \& Placido, 2010). Equivalently, a rate of 0.59 preferred act $f_1$ over act $f_2$, and a rate of 0.63 preferred act $f_4$ over $f_3$. As we have seen in Section \ref{ellsbergmachina}, this result is problematical also from the point of view of Choquet expected utility.

A quantum mechanical model for the experimental data above can be constructed by finding two orthogonal states $p_{w_1}$ and $p_{w_2}$, represented by the unit vectors $|w_1\rangle$ and $|w_2\rangle$, respectively, such that
\begin{eqnarray}
\langle w_1|\hat{F}_{1}-\hat{F}_{2}|w_1\rangle=0.59 \label{data1mach} \\
\langle w_2|\hat{F}_{4}-\hat{F}_{3}|w_2\rangle=0.63 \label{data2mach}
\end{eqnarray} 
where $\hat{F}_{i}$, $i=1,2,3,4$, are defined in (\ref{f1mach})--(\ref{f4mach}). In the canonical basis of ${\mathbb C}^{4}$, the solution is
\begin{eqnarray}
|w_1 \rangle&=&(0, 0.71 e^{i 1.6^{\circ}}, 0.38 e^{i 1^{\circ}}, 0.60 e^{i 185.2^{\circ}}) \label{w1machina}  \\
|w_2 \rangle&=&(0.71 e^{i 0.7^{\circ}},0.05 e^{i 191.8^{\circ}}, 0.62 e^{i 2.9^{\circ}}, 0.34 e^{i 7.4^{\circ}}) \label{w2machina}
\end{eqnarray}
as we have proved in Aerts \& Sozzo (2016). The states $p_{w_1}$ and $p_{w_2}$ represented in (\ref{w1machina}) and (\ref{w2machina}) identify the subjective probability distributions $\mu_{w_1}$ and $\mu_{w_2}$, respectively, reproducing the experimental pattern, for an utility value $u(50)-u(25)\approx 1.636$. But, generally speaking, preferences do depend on the state $p_v$ of the Machina entity. This completes the construction of a quantum model for the Machina reflection example with lower tail shifts, which follows the prescriptions in Section \ref{quantummodels}.

L'Haridon and Placido (2010) also tested the reflection example with upper tail shifts. The authors asked the same 94 students to rank the four acts  in Table 3, Section \ref{ellsbergmachina}. The students' response was that
8 students preferred acts $f_1$ and $f_3$, 
47 students preferred acts $f_1$ and $f_4$, 
6 students preferred acts $f_2$ and $f_4$, and
33 students preferred acts $f_2$ and $f_3$. This entails that $47+33=80$ students over 94 reversed their preferences, for a ratio of 0.85, thus violating subjective EUT and in agreement with Machina's expectations. Equivalently,  
a rate of 0.59 preferred act $f_1$ over act $f_2$, and a rate of 0.56 preferred act $f_4$ over $f_3$.

A quantum model for these experimental data can be constructed by following along the lines above. We have proved in 
Aerts \& Sozzo (2016) that the states $p_{w_1}$ and $p_{w_2}$ reproducing the data collected in the Machina reflection example with upper tail shifts are represented by the unit vectors
\begin{eqnarray}
|w_1 \rangle&=&(0.02 e^{i 0.3^{\circ}}, 0.71 e^{i 11.6^{\circ}}, 0.38 e^{i 1.3^{\circ}}, 0.60 e^{i 196.5^{\circ}})  \\
|w_2 \rangle&=&(0.71 e^{i 0.7^{\circ}},0, 0.59 e^{i 1.7^{\circ}}, 0.39 e^{i 16.9^{\circ}})
\end{eqnarray}
for an utility value $u(50)-u(25)=1.636$.

In all the examples above, we have proved that it is possible to reconstruct the quantum states, hence the quantum probability distributions representing the priors underlying concrete decisions in the presence of ambiguity. However, we have also argued that a more complete mathematical treatment suggests that preferences under ambiguity depend on the state of the DM entity, and that a DM entity should be associated with a family of subjective probability distributions parametrized by its states.

The modeling of the Machina paradox situations -- we have also modeled the `50:51 example' in Aerts, Sozzo \& Tapia (2012) -- is relevant, in our opinion, because the most established non-Bayesian models face difficulties when trying to reproduce the expected pattern (Machina, 2009; Baillon, L'Harison \& Placido, 2011).

\subsection{Possible extensions and a definition of ambiguity\label{quantumambiguity}}
We conclude the presentation of the quantum theoretical framework with some technical remarks.
 
The quantum theoretical model for human preferences can be extended to the case in which the set of states of nature ${\mathscr S}$  has a continuous cardinality, as follows.

We introduce, in addition to ${\mathscr S}$, the set of consequences ${\mathscr X}$, the set of acts ${\mathscr F}=\{ f:{\mathscr S} \longrightarrow {\mathscr X} \}$, the utility function $u:{\mathscr X}\longrightarrow \Re$. Further, let ${\cal H}$ be the Hilbert space representing states of the DM entity, and let $\{ |\alpha\rangle\}$ be an orthonormal basis of ${\cal H}$, so that $\langle \alpha|\alpha'\rangle=\delta(\alpha-\alpha')$, where $\delta(\cdot)$ is the $\delta$-Dirac distribution.

For every state $p_v \in \Sigma_{DM}$ describing the DM entity, the represented vector $|v\rangle$ can be written as
\begin{equation}
|v\rangle=\int_{\Re}\langle \alpha|v\rangle |\alpha\rangle d\alpha=\int_{\Re} c(\alpha) |\alpha\rangle d\alpha
\end{equation}
An event $E$ is represented by the orthogonal projection operator $P_{E}=\int_{E}|\alpha\rangle\langle \alpha| d\alpha$. Hence, the subjective probability that the event $E$ occurs when the DM entity is in the state $p_v$ is
\begin{equation}
\mu_{v}(E)=\langle v | \Big ( \int_{E}|\alpha\rangle\langle \alpha| d\alpha   \Big )| v\rangle=\int_{E}|\langle \alpha|v\rangle|^{2}d\alpha=\int_{\Re} d|| P_{\alpha}|v\rangle||^{2}
\end{equation}
where the integral is intended in the Lebesgue sense.

The act $f$ is instead represented by the Hermitian operator
\begin{equation}
\hat{F}=\int_{\Re}u(f(\alpha))|\alpha\rangle\langle\alpha| d \alpha
\end{equation}
Hence, the expected utility of the act $f$ in the state $p_v$ is
\begin{equation} \label{Wcontinuous}
W_{v}(f)=\langle v| \hat{F}|v\rangle=\int_{\Re} u(f(\alpha))|c(\alpha)|^{2}d\alpha=\int_{\Re} u(f(\alpha))d ||P_{\alpha}|v\rangle ||^{2}
\end{equation}
The right side of (\ref{Wcontinuous}) is also useful when one does not specifies the cardinality of the set of states. It is indeed sufficient to require that the integral is intended in the Lebesgue-Stieltjes sense.

The treatment of the Ellsberg and Machina paradox situations enables providing a general definition of ambiguity within the quantum theoretical framework, as follows.

We say that `an event $E$ is unambiguous' if the subjective probability $\mu_{v}(E)$ does not depend on the state $p_v$ of the DM entity. In the Ellsberg three-color example, the event $E_R$: ``a red ball will be drawn'' is indeed unambiguous in the sense specified here. Indeed, for every state $p_v$ of the Ellsberg entity, represented by the unit vector $|v\rangle \in {\mathbb C}^{3}$, $\mu_{v}(E_R)=\frac{1}{3}$.

Finally, we say that `an act $f$ is unambiguous' if the expected utility $W_v(f)$ does not depend on the state $p_v$ of the DM entity. Again in the Ellsberg three-color example, the act $f_4$ is indeed unambiguous in the sense specified above, since,  for every state $p_v$ of the Ellsberg entity, represented by the unit vector $|v\rangle \in {\mathbb C}^{3}$, $W_{v}(f_4)=\frac{1}{3}(u(0)+u(100))$.

\section{Conclusive remarks\label{conclusions}}
We have worked out in this paper a theoretical framework to represent preferences and decisions under uncertainty. This framework uses the mathematical formalism of quantum theory. In it, subjective probabilities are represented by families of quantum probability distributions, parametrized by the state of the DM entity under investigation. The interaction with the overall cognitive landscape, which includes the decision-maker's pondering about two acts, provokes a change of state of the DM entity, which enables modeling of 
the ambiguity aversion affecting Ellsberg- and Machina-like preferences. However, the present theoretical framework is flexible enough to reproduce different experimental patterns arising from different attitudes toward ambiguity. We have also stressed that the present approach allows modeling of beliefs, but in a generally non-Bayesian setting where a kind of `contextual risk' is present. Finally, the present approach suggests the development of a quantum-based subjective EUT with state-dependent preferences between acts.

In our opinion, this quantum theoretical framework can be successfully used to represent ambiguity-laden situations outside pure decision theory, in particular in strategic management decisions and some long standing economic puzzles. We conclude this paper with some hints in this direction.

Experiments have revealed a mixture of ambiguity aversion and ambiguity attraction in decisions under uncertainty. In this respect, managers typically compare the performance of an investment with a benchmark, or targeted performance, e.g., the return on investment (ROI), or the internal rate of return (IRR). A `gain' is realized when the performance is above the benchmark, a `loss' is realized when the performance is below the benchmark. For example, risk occurs in a situation where the probability that the ROI of a given investment is above the benchmark is $x$ per cent. Ambiguity occurs instead in a situation where the probability that the ROI of the investment is above the benchmark is between $(x - \Delta)$ and $(x+\Delta)$. Viscusi and Chesson (1999) identified  a `fear effect', as well as a `hope effect', in their experimental study. More precisely, they found that, as the probability of a loss increases, managers become less ambiguity averse, reaching a `crossover point' at which they become ambiguity seeking, which indicates a shift from a fear to a hope effect. Viceversa, as the probability of a gain increases, managers become less ambiguity seeking, reaching a `crossover point' at which they become ambiguity averse, which indicates this time a shift from a hope to a fear effect. This result was confirmed by Ho, Keller and Keltyka (2002). The quantum theoretical framework can reproduce this dual behavior by reconstructing the quantum states, hence the subjective probability distributions underlying the observed behavior, as we have done in the three color Ellsberg urn and the Machina reflection example (Aerts \& Sozzo, 2016).

Coming to economics and finance, Epstein and Miao (2003) have put forward that ambiguity aversion may explain the `home bias puzzle' in international finance: people prefer to trade stocks of their own country rather than foreign stocks. This is in principle compatible with the quantum theoretical approach, where ambiguity aversion can be explained in terms of the overall cognitive landscape surrounding the decision-making situation. Further, Hansen, Sargent and Wang (2002) have shown that ambiguity aversion models lead to market prices that are closer to the empirical prices than the prices predicted by EUT, which may explain the `equity premium puzzle'. Again, this behavior can in principle be reproduced within a theoretical framework where ambiguity aversion depends on the way the cognitive landscape influences the DM entity.

We believe that the quantum theoretical approach presented here can be naturally applied to the fascinating problems above, and we plan to dedicate future research to this.


\section*{References}
\begin{description}




\item Aerts, D. (2009). Quantum structure in cognition. {\it Journal of Mathematical Psychology 53}, 314--348.

\item Aerts, D., Broekaert, J., Gabora, L., \& Sozzo, S. (2013). Quantum structure and human thought. {Behavioral and Brain Sciences 36}, 274--276.



\item Aerts, D., Gabora, L., \& S. Sozzo, S. (2013). Concepts and their dynamics: A quantum--theoretic modeling of human thought. {\it Topics in Cognitive Science 5}, 737--772.


\item Aerts, D., Sozzo, S., \& Tapia, J. (2012) A quantum model for the Ellsberg and Machina paradoxes. {\it Quantum Interaction. Lecture Notes in Computer Science 7620}, 48--59.

\item Aerts, D., Sozzo, S., \& Tapia, J. (2014). Identifying quantum structures in the Ellsberg paradox. {\it International Journal of Theoretical Physics 53}, 3666--3682.


\item Aerts, D., Sozzo, S., \& Veloz, T. (2016). A new fundamental evidence of non-classical structure in the combination of natural concepts. \emph{Philosophical Transactions of the Royal Society A 374}, 20150095.

\item D. Aerts \& S. Sozzo (2016). From ambiguity aversion to a generalized expected utility. Modeling preferences in a quantum probabilistic framework. {\it Journal of Mathematical Psychology 74}, 117--127.

\item Allais, M. (1953). Le comportement de l'homme rationnel devant le risque. Critique des postulats et axiomes de l'ecole Am\'ericaine. {\it Econometrica 21}, 503--546.

\item Baillon, A., l'Haridon, O., \& Placido, L. (2011). Ambiguity models and the Machina paradoxes. {\it American Economic Review 101}, 1547--1560.

\item Binmore, K., Stewart, L., \& Voorhoeve, A. (2012). How much ambiguity aversion? Finding indifferences between Ellsberg’s risky and ambiguous bets. {\it Journal of Risk and Uncertainty 45}, 215--238.

\item Busemeyer, J. R., \& Bruza, P. D. (2012). {\it Quantum Models of Cognition and Decision}. Cambridge: Cambridge University Press.


\item Busemeyer, J. R., Pothos, E. M., Franco, R., \& Trueblood, J. S. (2011). A quantum theoretical explanation for probability judgment errors. {\it Psychological Review 118}, 193--218.

\item Camerer, C., \& Weber, M. (1992). Recent developments in modeling preferences: Uncertainty and ambiguity. {\it Journal of Risk and Uncertainty 5}, 325--370.

\item Dirac, P. A. M. (1958). {\it Quantum mechanics}, 4th ed. London: Oxford University Press.

\item Einhorn, H., \& Hogarth, R. (1986). Decision making under ambiguity. {\it Journal of Business 59} (Supplement), S225--S250.

\item Ellsberg, D.(1961). Risk, ambiguity, and the Savage axioms. {\it Quarterly Journal of Economics 75}. 643--669.

\item Epstein, L. G. (1999). A Definition of uncertainty aversion. {\it Review of Economic Studies 66}, 579--608.

\item Epstein, L. G. \& J. Miao (2003). A two-person dynamic equilibrium under ambiguity. {\it Journal of Economic Dynamics and Control, 27}, 1253--1288.

\item Fishburn, P. C. (1970). {\it Utility Theory for Decision Making}. New York: John Wiley and Sons.

\item Fox, C. R., \& Tversky, A. (1995). Ambiguity aversion and comparative ignorance. {\it The Quarterly Journal of Economics 110}, 585--603.


\item Ghirardato, P., Maccheroni, F., \&  Marinacci, M. (2004). Differentiating ambiguity and ambiguity attitude. {\it Journal of Economic Theory 118}, 133--173.

\item Gilboa, I., \& M. Marinacci (2013). Ambiguity and the Bayesian paradigm. In {\it Advances in Economics and Econometrics: Theory and Applications}.  Acemoglu, D., Arellano, M., \& Dekel, E. (Eds.)  New York: Cambridge University Press. 

\item Gilboa, I., A. Postlewaite, \& D. Schmeidler (2008). Probabilities in economic modeling. {\it Journal of Economic Perspectives, 22}, 173--188.

\item Gilboa, I., \& Schmeidler, D. (1989). Maxmin expected utility with a non-unique prior. {\it Journal of Mathematical Economics 18}, 141--153.

\item Gleason, A. M. (1957). Measures on the closed subspaces of a Hilbert space. {\it Indiana University Mathematics Journal 6}, 885--893.

\item Hansen, L., \& Sargent, T. (2001). Robust control and model uncertainty. {\it American Economic Review 91}, 60--66.

\item Hansen, L., Sargent, T., \& N. Wang, (2002). Robust permanent income and pricing with filtering. {\it Macroeconomic Dynamics, 6}, 40--84.

\item Haven, E., \& Khrennikov, A. Y. (2013). {\it Quantum Social Science}. Cambridge: Cambridge University Press.

\item Ho, J., Keller, L. R., \& Keltyka, P. (2002). Effects of outcome and probabilistic ambiguity on managerial choices. {\it Journal of Risk and Uncertainty 24}, 47--74.

\item Karni, E. (2014). Axiomatic foundations of expected utility and subjective probability. In Machina, M. J., \& Viscusi, K. (Eds.) {\it Handbook of the Economics of Risk and Uncertainty}, pp. 1--38. New York: Elsevier.

\item Khrennikov, A. Y. (2010). {\it Ubiquitous Quantum Structure}. Berlin: Springer. 

\item Khrennikov, A. Y. (2015). Quantum version of Aumann's approach to common knowledge: Sufficient conditions of impossibility to agree on disagree. {\it Journal of Mathematical Economics 60}, 89--104.

\item Klibanoff, P., Marinacci, M., \& Mukerji, S. (2005). A smooth model of decision making under ambiguity. {\it Econometrica 73}, 1849--1892.

\item Knight, F. H. (1921). {\it Risk, Uncertainty and Profit}. Boston: Houghton Mifflin.

\item Kolmogorov, A. N. (1933). {\it Grundbegriffe der Wahrscheinlichkeitrechnung}, Ergebnisse Der Mathematik; translated as {\it Foundations of Probability}. New York: Chelsea Publishing Company, 1950.

\item L'Haridon, O., \& Placido, L. (2010). Betting on Machina’s reflection example: An experiment on ambiguity. {\it Theory and Decision 69}, 375--393.


\item Machina, M. J. (2009). Risk, ambiguity, and the dark–dependence axioms. {\it American Economical Review 99}. 385–-392.

\item Machina, M. J., \& Siniscalchi, M. (2014). Ambiguity and ambiguity aversion. In Machina, M. J., \& Viscusi, K. (Eds.) {\it Handbook of the Economics of Risk and Uncertainty}, pp. 729--807. New York: Elsevier.

\item Maccheroni, F., Marinacci, M., \& Rustichini, A. (2006a). Ambiguity aversion, robustness, and the variational representation of preferences. {\it Econometrica 74}, 1447--1498.

\item McCrimmon, K., \& Larsson, S. (1979). Utility theory: Axioms versus paradoxes. In Allais, M., \& Hagen, O., (Eds.) {\it Expected Utility hHypotheses and the Allais Paradox}, pp. 27--145. Dordrecht: Reidel.

\item Pitowsky, I. (1989). {\it Quantum Probability, Quantum Logic}. Lecture Notes in Physics vol. {\bf 321}.  Berlin: Springer.

\item Pothos, E. M., \& Busemeyer, J. R. (2009). A quantum probability explanation for violations of `rational' decision theory. {\it Proceedings of the Royal Society B 276}, 2171--2178.

\item Pothos, E. M., \& Busemeyer, J. R. (2013). Can quantum probability provide a new direction for cognitive modeling? {\it Behavioral and Brain Sciences 36}. 255--274.

\item Savage, L. (1954). {\it The Foundations of Statistics}. New York: John Wiley \& Sons. Revised and Enlarged Edition (1972), New York: Dover Publications.

\item Schmeidler, D. (1989). Subjective probability and expected utility without additivity. {\it Econometrica 57}, 571--587.

\item Slovic, P., \& Tversky, A. (1974). Who accepts Savage' s axiom? {\it Behavioral Science 19}, 368--373.

\item Sozzo, S. (2015). Conjunction and negation of natural concepts: A quantum-theoretic framework. {\it Journal of Mathematical Psychology 66}, 83--102.

\item Tversky, A., \& Kahneman, D. (1974). Judgment under uncertainty: Heuristics and biases. {\it Science 185}, 1124--1131.

\item Tversky, A. \& Kahneman, D. (1983). Extension versus intuitive reasoning: The conjunction fallacy in probability judgment. {\it Psychological Review 90}, 293--315.

\item Tversky, A., \& Kahneman, D. (1992). Advances in prospect theory: Cumulative representation of uncertainty. {\it Journal of Risk and Uncertainty 5}, 297--323.

\item Tversky, A., \& Shafir, E. (1992). The disjunction effect in choice under uncertainty. {\it Psychological Science 3}, 305--309.

\item Viscusi, W. K., \& Chesson, H. (1999). Hopes and fears: The conflicting effects of risk ambiguity. {\it Journal of Risk and Uncertainty 47}, 153--178.

\item von Neumann, J., \& Morgenstern, O. (1944). {\it Theory of Games and Economic Behavior}. Princeton: Princeton University Press.


\item Wang, Z., Solloway, T., Shiffrin, R. M., Busemeyer, J. R. (2014). Context effects produced by question orders reveal quantum nature of human judgments. {\it Proceedings of the National Academy of Sciences 111}, 9431--9436.

\end{description}

\end{document}